\documentclass[11pt]{article}
\usepackage{amsmath,amssymb,color,epsfig,cite}
\usepackage{graphicx}
\usepackage{subfigure}
\usepackage{setspace}

\usepackage{amsfonts}

\newcommand{\hoch}[1]{$\, ^{#1}$}

\makeatletter
\@addtoreset{equation}{section}
\makeatother

\newcommand{\be}{\begin{equation}}
\newcommand{\ee}{\end{equation}}
\newcommand{\bea}{\setlength\arraycolsep{2pt} \begin{eqnarray}}
\newcommand{\eea}{\end{eqnarray}}
\newcommand{\nn}{\nonumber}

\def\ft#1#2{{\textstyle{\frac{\scriptstyle #1}{\scriptstyle #2} } }}
\def\fft#1#2{{\frac{#1}{#2}}}

\def\0{{\sst{(0)}}}
\def\1{{\sst{(1)}}}
\def\2{{\sst{(2)}}}
\def\3{{\sst{(3)}}}
\def\4{{\sst{(4)}}}
\def\5{{\sst{(5)}}}
\def\6{{\sst{(6)}}}
\def\7{{\sst{(7)}}}
\def\8{{\sst{(8)}}}
\def\9{{\sst{(9)}}}
\def\smallten{{\sst{(10)}}}
\def\sst#1{{\scriptscriptstyle #1}}

\begin{document}



\begin{center}
{\large {\bf Quasi-Topological Ricci Polynomial Gravities}}

\vspace{10pt}
Yue-Zhou Li\hoch{1\dagger}, Hai-Shan Liu\hoch{2\ddagger} and  H. L\"u\hoch{1*}

\vspace{15pt}

\hoch{1}{\it Department of Physics, Tianjin University, Tianjin 300072, China}

\vspace{10pt}

\hoch{2}{\it Institute for Advanced Physics \& Mathematics,\\
Zhejiang University of Technology, Hangzhou 310023, China}

\vspace{30pt}

\underline{ABSTRACT}

\end{center}

Quasi-topological terms in gravity can be viewed as those that give no contribution to the equations of motion for a special subclass of metric ans\"atze.  They therefore play no r\^ole in constructing these solutions, but can affect the general perturbations. We consider Einstein gravity extended with Ricci tensor polynomial invariants, which admits Einstein metrics with appropriate effective cosmological constants as its vacuum solutions.  We construct three types of quasi-topological gravities.  The first type is for the most general static metrics with spherical, toroidal or hyperbolic isometries.  The second type is for the special static metrics where $g_{tt} g_{rr}$ is constant.  The third type is the linearized quasi-topological gravities on the Einstein metrics. We construct and classify results that are either dependent on or independent of dimensions, up to the tenth order. We then consider a subset of these three types and obtain Lovelock-like quasi-topological gravities, that are independent of the dimensions.  The linearized gravities on Einstein metrics on all dimensions are simply Einstein and hence ghost free.  The theories become quasi-topological on static metrics in one specific dimension, but non-trivial in others.  We also focus on the quasi-topological Ricci cubic invariant in four dimensions as a specific example to study its effect on holography, including shear viscosity, thermoelectric DC conductivities and butterfly velocity.  In particular, we find that the holographic diffusivity bounds can be violated by the quasi-topological terms, which can induce an extra massive mode that yields a butterfly velocity unbound above.

\vfill {\footnotesize \hoch{\dagger}liyuezhou@tju.edu.cn\ \ \ \hoch{\ddagger} hsliu.zju@gmail.com\ \ \ \hoch{*}mrhonglu@gmail.com}

\pagebreak

\tableofcontents
\addtocontents{toc}{\protect\setcounter{tocdepth}{2}}


\newpage

\section{Introduction}

Einstein gravity extended with quadratic curvature invariants is the first step to include an infinite number of possible higher-order curvature terms.  In four dimensions, owing to the fact that the Gauss-Bonnet combination $\sqrt{-g} (R^2 -4 R_{\mu\nu}R^{\mu\nu} + R_{\mu\nu\rho\sigma} R^{\mu\nu\rho\sigma})$ is a total derivative, quadratic terms involve only the Ricci tensor polynomials.  The theory has attracted continued attention since it was proven that it is generally renormalizable, albeit it contains additional ghost-like massive spin-2 modes \cite{Stelle:1976gc,Stelle:1977ry}.  The theory admits the usual Schwarzschild black hole as a vacuum solution.  Recently it was demonstrated numerically that the theory contains a new black hole associated with the condensation of the massive spin-2 modes \cite{Lu:2015cqa,Lu:2015psa}. The result was generalized to higher dimensions \cite{Lu:2017kzi}.  (See, e.g.~\cite{Boulware:1985wk,Deser:2001pe,Deser:2002rt,Li:2008dq,Bergshoeff:2009hq,
Liu:2009bk,Lu:2011zk,Deser:2011xc,Liu:2011kf}, for studies on linearized theories of quadratically-extended gravities.)

Einstein gravity extended with higher-order Riemann curvature polynomial invariants in general admits multiple vacua that are maximally symmetric spacetimes, such as the Minkowski, de Siter (dS) or anti-de Sitter (AdS) spacetimes.  Linearized gravity around these vacua typically involves quartic differential equations, containing the usual massless graviton, as well as a possible massive scalar and a ghost-like massive (spin-2) graviton.  In some particular combinations, such as the Gauss-Bonnet or more general Lovelock terms \cite{ll}, the massive modes are absent and theory can be ghost free.  Interestingly, the $k$'th order Lovelock term vanishes in $D\le 2k-1$, and it is purely topological in $D=2k$ dimensions in that it is a total derivative and gives no contribution to the equations of motion.

In general relativity, or its application in gauge/gravity duality, black hole solutions are of particular importance.  Many insights can be gained from studying the static black holes with spherical, toroidal or hyperbolic isometries.  The most general ansatz for such metrics in $D$ dimensions is given by
\be
ds_D^2 = - h(r) dt^2 + \fft{dr^2}{f(r)} + r^2 d\Omega_{D-2,\epsilon}^2\,,\label{genmetric}
\ee
where $d\Omega_{D-2,\epsilon}^2$ is the metric of round $S^{D-2}$, $T^{D-2}$ and $H^{D-2}$ for $\epsilon=1,0,-1$ respectively.  Although analytical solutions of black holes in Einstein-Gauss-Bonnet gravity exist \cite{Boulware:1985wk,Cai:2001dz}, an analytical black hole in theories extended with higher-order curvature invariants are hard to come by. The existence of additional ghost modes implies that such a black hole is unlikely to be unique, as demonstrated in \cite{Lu:2015cqa}.  It is thus of interest to construct curvature polynomials that give no contribution to the equations of motion for the ansatz (\ref{genmetric}).  The $k$'th-order Lovelock term in $D=2k$ dimensions fits the criterium, but it is topological and it has no effect on constructing any metrics except for measuring their topological characteristics.

The quasi-topological structure on the other hand is more interesting: it has no effect on the equations of motion for the static metric (\ref{genmetric}), but it can have nontrivial effects on the general perturbations.  This is a compromise that leaves the background solutions intact but alters the perturbative spectrum.  Quasi-topological Riemann cubic gravity in six dimensions were constructed in \cite{Myers:2010ru}.  Unlike the third-order Lovelock term that vanishes in five dimensions, the quasi-topological cubic theory is non-vanishing, and its holographic properties were studied \cite{Myers:2010jv}. The work has attracted much attention and here is the list of incomplete references \cite{Brenna:2011gp,Oliva:2011xu,Dehghani:2011hm,Dehghani:2011vu,Oliva:2012zs,
Brenna:2012gp,Bazrafshan:2012rn,
Dehghani:2013ldu,Ghanaatian:2014bpa,Ghanaatian:2015qra,Bueno:2016xff,Hennigar:2016gkm,Bueno:2016lrh,
Chernicoff:2016qrc,
Cisterna:2017umf,Hennigar:2017ego,Dykaar:2017mba,Bueno:2017sui,Ahmed:2017jod,Bueno:2017qce,Feng:2017tev}.  A new notable example is a quartic polynomial that is quasi-topological in eight dimensions \cite{Dehghani:2011vu}.

Quasi-topological gravity was perhaps a serendipitous discovery. In literature, the construction typically starts from general Riemann tensor polynomials and then requires that the solution space contains the Schwarzschild-like black hole with $h=f$, and furthermore, there exists a first integration.  (See, e.g.~\cite{Oliva:2010eb,Myers:2010ru}.) It turns out that in some cases the resulting theory becomes quasi-topological in certain specific dimensions, like the cubic polynomial in six \cite{Myers:2010jv} and the quartic polynomial in eight dimensions \cite{Dehghani:2011vu}.  However, since the polynomial coefficients typically depend on the dimension, the cubic and quartic theories in six and eight dimensions do not have any special properties in other dimensions for the same set of coefficients.

In this paper, we focus our attention purely on the Ricci tensor polynomial invariants.  These polynomials are much more manageable compared to those constructed from the general Riemann tensor.  We study these structures up to the tenth order, since at the tenth order, the total number of the polynomials is exactly 42! An important benefit of Ricci gravity is that the Schwarzschild black hole ($h=f$) is already a solution.  In fact all Einstein metrics with appropriate cosmological constants are solutions of the theory.  This leads to another benefit that one can construct covariant linearized gravity on the Einstein metrics.  By contrast, the covariant linearized theory of a generic Riemann tensor polynomial is possible only on a maximally-symmetric vacuum. We seek for combinations for which there is neither massive scalar nor ghost-like massive spin-2 mode in linearized gravity around the Einstein metrics.  We expect that this ensures that the Schwarzschild metric is the only black hole solution out of the ansatz (\ref{genmetric}).  We then look for combinations such that the Ricci polynomials give no contribution to the equations of motion for the ansatz (\ref{genmetric}).  We find that the first occurrence is at the quintic order in four dimensions, and at the sextic order in $D\ne 4$ dimensions.  We call these quasi-topological Ricci polynomial gravities on general static metrics.  Thus our construction is different from those in literature.  By construction, all our polynomials are quasi-topological and give no contribution to the equations of motion.

For large classes of solutions known in literature, including Schwarzschild and Reissner-Nordstr\o m (RN) black holes, the metric functions $h$ and $f$ are equal, leading to a special class of static black hole metrics
\be
ds_D^2 = - f(r) dt^2 + \fft{dr^2}{f(r)} + r^2 d\Omega_{D-2,\epsilon}^2\,.\label{spemetric}
\ee
In other words, the product $g_{tt} g_{rr}$ is constant. We can also construct polynomials such that they give no contribution to the equations of motion for this special class of static metrics.  It is clear that the restrictions in this case are more relaxed and consequently the lowest example is at the cubic order in four dimensions, and at the quartic order in general $D\ne 4$ dimensions.  We call these quasi-topological Ricci polynomials on special static metrics.

As being mentioned earlier, Ricci gravities admit Einstein metrics as vacuum solutions.  We find that the absence of massive scalar and spin-2 modes in linearized Ricci gravities on Einstein metrics implies that the higher-order terms give no contribution to these linearized theories either, as if they were purely topological. The resulting theories are exactly the same as linearized Einstein gravity and therefore are ghost free at the linear level.  We call these linearized quasi-topological gravities.  The aforementioned  quasi-topological gravities on both general and special static metrics are subset classes of these linearized quasi-topological theories, and hence they are all ghost free.

The coefficients of the quasi-topological gravities are in general dependent on the dimensions $D$, and hence they are not the same theories in different dimensions.  We find further subclasses of dimension independent polynomials, in which case, the same theory is quasi-topological in all dimensions.  Furthermore, it is of interest to construct Lovelock-like theories that are dimension independent and are quasi-topological in some specific dimensions, but nontrivial and ghost free in other dimensions.  We find that such Ricci gravities do exist.

It is worth clarifying here that our construction becomes degenerate in one and two dimensions.  When we refer in this paper ``general'' or ``all'' dimensions, it is understood that they mean all dimensions $D\ge 3$.  Also in this paper, when we say a theory is ghost free, we mean only that the linearized gravity on an Einstein metric, rather than on a generic metric, contains no ghost excitations.  Finally, when we use the terminology of ``general'' static ansatz, we mean (\ref{genmetric}), with spherical, toroidal or hyperbolic isometries.  The ``special'' static metric ansatz refers to (\ref{spemetric}).

      One motivation for us to construct these quasi-topological gravities is to
study the effect of higher-order terms on the AdS/CFT correspondence \cite{Maldacena:1997re}, in subjects related to strongly-coupled condensed matter physics in particular \cite{ggd1,ggd2,ggd3,ggd4}.
Our construction allows that all the previously known static AdS black holes continue to be solutions when our quasi-topological terms are included in the theory.  This significantly simplifies the task of constructing the background solutions in the presence of higher-order curvature terms. On the other hand, the linear perturbations can be affected by these higher-order terms, which are not topological, but only quasi-topological. In this paper, we employ a concrete example of quasi-topological Einstein-Maxwell Ricci cubic gravity in four dimensions, and study the corresponding holographic shear viscosity, thermoelectric conductivities and also butterfly velocity. In particular, we find that conjectured holographic diffusivity bounds \cite{elebound,thebound} can be violated by the quasi-topological terms, not by changing the thermoelectric properties, but by inducing a new massive mode that yields a butterfly velocity that is unbound above.

The paper is organized as follows.  In section 2, we study Einstein gravity extended with cubic Ricci tensor invariants.  We derive the equations of motion and obtain the effective cosmological constants of the Einstein metric solutions.  We construct linearized gravity on an Einstein metric and require that the massive scalar and spin-2 modes be absent.  We find that the resulting linearized gravity is exactly the same as linearized Einstein gravity, giving rise to linearized quasi-topological gravity.
Interestingly, we find that the theory becomes also quasi-topological on special static metrics of (\ref{spemetric}) in four and only four dimensions.  We also argue that Schwarzschild black holes are only solutions out of the ansatz (\ref{genmetric}) in theories we construct in this paper. In section 3, we generalize the discussion up to the tenth order of Ricci polynomials.  We construct quasi-topological Ricci gravities for both general and special static metrics, as well as linearized quasi-topological gravities.  We classify the results that are either dependent on or independent of dimensions, and also separate the irreducible solutions from the reducible ones.
In section 4, we construct Lovelock-like quasi-topological Ricci gravities, which are nontrivial and ghost free in all but one specific dimension.  In section 5, we consider the Einstein-Maxwell-Axion theory extended with the quasi-topological Ricci cubic term in four dimensions and study the AdS/CFT correspondence.  We summarize our results and conclude the paper in section 6.  In appendix A, we present the full results of irreducible quasi-topological Ricci polynomials we construct in this paper.  The complete structure of quasi-topological Ricci gravities can be obtained from these irreducible solutions.  In appendix B, we present a proof of the matching of the radially-conserved bulk current with boundary stress tensor for general Ricci cubic gravity.  This matching is important for us to read off holographic properties from the bulk currents in the study of the AdS/CFT correspondence in section 5.

\section{Quasi-topological Ricci cubic gravity}

\label{sec:cubic}

\subsection{Einstein gravity with cubic Ricci polynomials}

We begin in this section to consider Einstein gravity extended with cubic Ricci-tensor polynomial invariants.  The bulk action in general $D$ dimensions is given by
\be
S_{\rm bulk} = \fft{1}{16\pi}\int d^{D}x\, \sqrt{-g}\, L\,,
\ee
where $L=L_0 + L_1 + L_2 + L_3$, with
\be
L_0 = \kappa_0 (R - 2\Lambda_0)\,,\qquad
L_1 = e_1 R^3\,,\qquad L_2 = e_2 R\,R_{\mu\nu} R^{\mu\nu}\,,\qquad
L_3= e_3 R^{\mu}_{\nu} R^{\nu}_\rho R^{\rho}_\mu\,.
\ee
Note that we introduced $\kappa_0$, the inverse of Newton's constant, the bare cosmological constant $\Lambda_0$ and three coupling constants $(e_1,e_2,e_3)$ associated with the three cubic invariants. The covariant equation of motion associated with the variation of the metric $\delta g^{\mu\nu}$ is
\be
{\cal E}_{\mu\nu} \equiv  P_{\mu \alpha\beta\gamma}R_\nu{}^{\alpha\beta\gamma}-\ft12g_{\mu\nu}{L}-2\nabla^\alpha\nabla^\beta P_{\mu\alpha\beta\nu} = 0\,,\label{geneom}
\ee
where
\be
P_{\mu\nu\rho\sigma} \equiv \fft{\partial L}{\partial R^{\mu\nu\rho\sigma}}\,.
\ee
For the $(L_0,L_1,L_2,L_3)$ terms in the Lagrangian, the corresponding $P$ tensors are given by
\bea
P^0_{\mu\nu\rho\sigma} &=& \ft12\kappa_0 (g_{\mu\rho} g_{\nu\sigma} - g_{\mu\sigma} g_{\nu\rho})\,,\qquad
P^1_{\mu\nu\rho\sigma} = \ft32e_1 R^2  (g_{\mu\rho} g_{\nu\sigma} - g_{\mu\sigma} g_{\nu\rho})\,,\cr
P^2_{\mu\nu\rho\sigma} &=& \ft12e_2 R_{\alpha\beta}R^{\alpha\beta} (g_{\mu\rho} g_{\nu\sigma} - g_{\mu\sigma} g_{\nu\rho}) \cr
&& + \ft12e_2 R\, (g_{\mu\rho} R_{\nu\sigma} - g_{\mu\sigma} R_{\nu\rho}
-g_{\nu\rho} R_{\mu\sigma} + g_{\nu\sigma} R_{\mu\rho})\,,\cr
P^3_{\mu\nu\rho\sigma} &=& \ft34 e_3 \big(g_{\mu\rho} R_{\nu\gamma}R_{\sigma}{}^\gamma - g_{\mu\sigma} R_{\nu\gamma} R_{\rho}{}^\gamma  -
g_{\nu\rho} R_{\mu\gamma}R_{\sigma}{}^\gamma + g_{\nu\sigma} R_{\mu\gamma} R_{\rho}{}^\gamma\big)
\eea
It is clear that the theory admits Einstein metrics as vacuum solutions.  It is convenient to define an effective cosmological constant $\Lambda_{\rm eff}$, namely
\be
\bar R_{\mu\nu} = \fft{2\Lambda_{\rm eff}}{D-2} \bar g_{\mu\nu}\equiv a \bar g_{\mu\nu}\,.
\ee
In this convention, for negative $\Lambda_{\rm eff}$, the radius $\ell$ of the AdS vacuum
is
\be
\Lambda_{\rm eff}=-\fft{(D-1)(D-2)}{2\ell^2}\,.
\ee
For this ansatz, we have
\bea
\bar P^0_{\mu\nu\rho\sigma} &=& \ft12\kappa_0 (\bar g_{\mu\rho} \bar g_{\nu\sigma} - \bar g_{\mu\sigma} \bar g_{\nu\rho})\,,\qquad
\bar P^1_{\mu\nu\rho\sigma} = \ft32 D^2 a^2 e_1 (\bar g_{\mu\rho} \bar g_{\nu\sigma} - \bar g_{\mu\sigma} \bar g_{\nu\rho})\,,\nn\\
\bar P^2_{\mu\nu\rho\sigma} &=& \ft32 D a^2 e_2  (\bar g_{\mu\rho} \bar g_{\nu\sigma} - \bar g_{\mu\sigma} \bar g_{\nu\rho})\,,\qquad
\bar P^3_{\mu\nu\rho\sigma} = \ft32 a^2 e_3 (\bar g_{\mu\rho} \bar g_{\nu\sigma} - \bar g_{\mu\sigma} \bar g_{\nu\rho})\,.
\eea
Substituting these into the equations of motion (\ref{geneom}), we have
\bea
\kappa_0 \Lambda_0 &=& \ft12\kappa_0(D-2) a + \ft12(D-6) (D^2 e_1 + D e_2 + e_3) a^3\nn\\
&=&\kappa_0 \Lambda_{\rm eff} + \ft{4(D-6)}{(D-2)^3} (D^2 e_1 + D e_2 + e_3) \Lambda_{\rm eff}^3\,.
\eea

\subsection{Linearized gravity and quasi-topology}

Having obtained the Einstein metrics as the vacuum solutions, we would like to perform linear perturbations on these backgrounds.  For $g_{\mu\nu}=\bar g_{\mu\nu} + \delta g_{\mu\nu}$, we define two quantities
\be
\delta \widetilde R_{\mu\nu} = \delta R_{\mu\nu} - a \delta g_{\mu\nu}\,,\qquad
\delta (R) = \delta (g^{\mu\nu} R_{\mu\nu}) = \bar g^{\mu\nu} \delta \widetilde R_{\mu\nu}\,.
\ee
After some somewhat involved algebra, we find each contribution to the linearized equations of motion associated with $\sqrt{-g} L_i$:
\bea
\delta {\cal E}_{\mu\nu}^0 &=& \kappa_0 \Big[\delta \widetilde R_{\mu\nu} - \ft12 \bar g_{\mu\nu} \delta(R) + \big(\Lambda_0 - \ft12(D-2)a\big)\delta g_{\mu\nu}\Big]\,,\cr
\delta {\cal E}_{\mu\nu}^1 &=& e_1\Big[3(Da)^2 \delta \widetilde R_{\mu\nu} -
\ft32 D(D-4) a^2 \bar g_{\mu\nu} \delta (R) - 6Da (\bar \nabla_\mu\bar \nabla_\nu -
\bar g_{\mu\nu} \bar \Box) \delta (R)\cr
&& - \ft12 D^2 a^3 (D-6) \delta g_{\mu\nu}\Big]\,,\cr
\delta {\cal E}_{\mu\nu}^2 &=& e_2\Big[-Da \bar \Delta_L \delta \widetilde R_{\mu\nu} -
(D+4)a \big(\bar\nabla_\mu \bar\nabla_\nu - \bar g_{\mu\nu} \bar \Box\big) \delta(R) -
\ft12 Da \bar g_{\mu\nu} \bar\Box \delta(R)\cr
&& + 5D a^2 \delta \widetilde R_{\mu\nu} - \ft12 (3D-8) a^2 \delta(R) -
\ft12 Da^3(D-6) \delta g_{\mu\nu}\Big]\,,\cr
\delta {\cal E}_{\mu\nu}^3 &=& e_3\Big[-3Da\bar \Delta_L \delta\widetilde R_{\mu\nu} -3a
\big(\bar\nabla_\mu \bar\nabla_\nu - \bar g_{\mu\nu} \bar \Box\big) \delta(R) -\ft32
a \bar g_{\mu\nu} \bar\Box \delta(R) + 9 a^2 \delta \widetilde R_{\mu\nu}\cr
&& -\ft32 a^2 \bar g_{\mu\nu} \delta(R) - \ft12 (D-6) a^3 \delta g_{\mu\nu}\Big]\,.
\eea
The total linearized equations therefore involve only $\delta \widetilde R_{\mu\nu}$ and
$\delta (R)$, given by
\bea
\delta {\cal E}_{\mu\nu}^{\rm tot} &=&\kappa_0\Big(\delta \widetilde R_{\mu\nu} - \ft12\bar g_{\mu\nu} \delta(R)\Big) + a^2(3D^2 e_1 + 5 D e_2 + 9 e_3) \delta\widetilde R_{\mu\nu} \nn\\
&&-\ft12 a^2 \big(3D(D-4) e_1 + (3D-8) e_2 + 3 e_3\big)\bar g_{\mu\nu} \delta(R)
\nn\\
&& -a (De_2 + 3 e_3) \bar \Delta_L \left( \delta \widetilde R_{\mu\nu} - \ft12 g_{\mu\nu} \delta(R)\right)\nn\\
&& -a \big(6De_1 + (D+4) e_2 + 3 e_3\big) \big(\bar\nabla_\mu \bar\nabla_\nu - \bar g_{\mu\nu} \bar \Box\big) \delta(R)=0\,.\label{cubiclineareom}
\eea
Here $\bar \Delta_L$ is the background Lichnerowic operator, given by
\bea
\bar \Delta_L \delta (R) &=& -\bar \Box\delta(R)\,,\cr
\bar \Delta_L \delta {\widetilde R}_{\mu\nu} &=& -\bar \Box \delta {\widetilde R}_{\mu\nu} - 2\bar R_\mu{}^\rho{}_\nu{}^\sigma \delta {\widetilde R}_{\rho\sigma} +
\bar R_{\mu}{}^{\rho} \delta {\widetilde R}_{\rho\nu} + \bar R_{\nu}{}^\rho\delta
{\widetilde R}_{\mu\rho}\,.
\eea
Linearized gravity of quadratic Ricci invariants was studied in \cite{Liu:2011kf,Lu:2017kzi}.

We now require that the linearized spectrum contains only the usual graviton mode, with no massive scalar or massive spin-2 modes.  This can be achieved by setting
\be
De_2 + 3e_3=0\,,\qquad 6 D e_1 + (D+4) e_2 + 3 e_3=0\,.\label{e1e2e3cons}
\ee
This leads to just one cubic Ricci tensor combination
\be
{\cal W}^\3=R^3 - \ft32D\, R\,R_{\mu\nu} R^{\mu\nu} + \ft12 D^2\, R^{\mu}_{\nu} R^{\nu}_\rho R^{\rho}_\mu\,.\label{P31}
\ee
Intriguingly, for this particular combination, the linearized gravity becomes simply that of Einstein theory, namely
\be
\delta {\cal E}_{\mu\nu}^{\rm tot} =\kappa_0\Big(\delta \widetilde R_{\mu\nu} - \ft12\bar g_{\mu\nu} \delta(R)\Big)\,.
\ee
In other words, the Ricci cubic term ${\cal W}^\3$ gives no contribution at all to the linearized theory, as if it is purely topological. However, it is clearly not topological for a generic metric at the nonlinear level.  It is therefore appropriate to call it linearized quasi-topological gravity on Einstein metrics.  This should be contrasted with Einstein-Gauss-Bonnet gravity in $D\ge 5$ or Einsteinian cubic gravity \cite{Bueno:2016xff} where the effective Newton's constant $1/\kappa_0$ is modified by the higher-order terms. (See also, \cite{Hennigar:2016gkm,Bueno:2016lrh,Feng:2017tev}.)

It is also instructive to study the linearized spectrum before we set the conditions (\ref{e1e2e3cons}).  The linearized equations (\ref{cubiclineareom}) can be split into the traceless and trace parts, namely
\bea
&&-a\alpha \Big(\bar \Delta_L - 2a + \mu_2^2\Big) \delta \hat R_{\mu\nu} =
a (\alpha + 2\beta) \big(\bar \nabla_\mu \bar \nabla_\nu  -
\fft{1}{D} \bar g_{\mu\nu}\bar \Box\big) \delta (R)\,,\nn\\
&&\ft12 a \big(D\alpha + 4(D-1) \beta\big) \Big(\bar \Box -\mu_0^2\Big) \delta (R) = 0\,,\qquad
\delta \hat R_{\mu\nu} = \delta \tilde R_{\mu\nu} - \fft1{D} \bar g_{\mu\nu} \delta (R)\,,\label{massivemodes}
\eea
where
\be
\alpha \equiv D e_2 + 3 e_3\,,\qquad \beta \equiv 3D e_1 + 2 e_2\,,
\ee
and the masses of the massive scalar and spin-2 modes are
\be
\mu_0^2 = \fft{D+2 + (D-6)(\alpha + D\beta)a^2}{(D\alpha + 4(D-1)\beta)a}\,,\qquad
\mu_2^2 = 2a - \fft{3(3\alpha + D\beta)a^2 +1}{\alpha a}\,.
\ee
Thus the decoupling of the massive modes requires either $a=0$ or $\alpha=0=\beta$, which corresponds to (\ref{e1e2e3cons}) and gives rise to (\ref{P31}).  The case with $a=0$, on the other hand, yields Ricci-flat spacetimes on which cubic curvature tensor indeed gives no contribution to the linearized equations of motion.  In this paper, we consider only the situation with $a\ne 0$.  One may also consider other interesting limits.  Critical gravity of \cite{Lu:2011zk,Deser:2011xc} can be achieved by setting $\mu_2=0$, together with the decoupling of the massive scalar modes by setting  $D\alpha + 4(D-1)\beta=0$.  The ghost free condition may be also possible if we decouple the massive spin-2 mode by setting $\alpha=0$. It was also demonstrated \cite{Lu:2017kzi} that the existence of the massive spin-2 mode as in (\ref{massivemodes}) implies that the theory admits a new branch of static black holes over and above the Schwarzschild one.  We shall come back to this point in section \ref{sec:uniqueness}.

It is worth emphasizing that covariant linearized gravities on Einstein metrics are luxuries enjoyed only by Ricci gravities.  In Riemann tensor gravities, covariant linearized gravity is possible only in Minkowski or (A)dS maximally symmetric vacua, unless the Riemann structure is purely topological.  There are a total of eight Riemann cubic invariants.  Absence of massive scalar and spin-2 modes also give rise to two conditions, leading to six Einsteinian linearized gravity.  The effective Newton's constant of these linearized theories is typically modified by the cubic terms.  Two examples are worth noting.  The first is the well-known third-order Lovelock combination. The other is the recently constructed Einsteinian cubic gravity \cite{Bueno:2016xff}.  The coefficients of both examples are independent of dimensions.

     The linearized quasi-topological Ricci cubic gravity (\ref{P31}) on the other hand depends on
the dimensions, and hence they are different theories in different dimensions.  Furthermore, the linearized theory with coefficients specified by parameter $D$ is ghost free only in $D$ dimensions, but not so in dimensions other than $D$.  In later sections, we shall construct quasi-topological gravities at higher orders that are independent of dimensions and ghost free at the linear level in all dimensions.

\subsection{Quasi-topological Ricci cubic gravity in $D=4$}

In four dimensions, something new happens to the cubic term (\ref{P31}).  We give a new name specifically for $D=4$:
\be
{\cal Q}^\3=R^3 - 6R\,R_{\mu\nu} R^{\mu\nu} + 8 R^{\mu}_{\nu} R^{\nu}_\rho R^{\rho}_\mu\,.\label{Q31}
\ee
If we substitute the special static metric ansatz (\ref{spemetric}), the equations of motion derived from the variation of $\sqrt{-g} {\cal Q}^\3$ is automatically satisfied, as if it were topological.  It is not the case for the general static metric (\ref{genmetric}) with $h\ne f$.  Thus we have constructed four dimensional quasi-topological Ricci cubic gravity on
special static metrics. (Note that the Ricci cubic structure ${\cal Q}^\3$ was also obtained in
\cite{Hennigar:2017ego} in somewhat different context.)

      There are a large class of static black hole solutions in four dimensions with $h=f$,
including the celebrated RN black hole.  These black holes continue to be solutions after ${\cal Q}^\3$ is included in the theory.  However, linearized gravity in these black hole backgrounds will involve nontrivial contributions from ${\cal Q}^\3$.  In section \ref{sec:AdS/CFT}, we shall discuss how this phenomenon affects the linear response system in the AdS/CFT correspondence.

\subsection{On the uniqueness of Schwarzschild black hole}
\label{sec:uniqueness}

In Ricci polynomial gravities, Einstein metrics with appropriate cosmological constant are solutions of the theories.  The static solution with spherical symmetry is the celebrated Schwarzschild black hole.  As was demonstrated in \cite{Lu:2015cqa}, new black holes associated with the condensation of massive modes can emerge.  As a concrete and simple example, we consider Einstein gravity extended with quadratic curvature invariants in four dimensions, studied in \cite{Stelle:1976gc,Stelle:1977ry}, namely
\be
L=R + \alpha R^2 +\beta R_{\mu\nu} R^{\mu\nu}\,.
\ee
Minkowski spacetime is the vacuum of the theory. Following from the metric ansatz (\ref{genmetric}), the static and spherically symmetric deviation from the vacuum at the linear order is \cite{Stelle:1976gc,Stelle:1977ry}
\bea
h &=& 1  - \fft{m}{r}  - \fft{c_0}{r} e^{-\mu_0 r} - \fft{c_2}{r} e^{-\mu_2 r} + \cdots
\,,\cr
f &=& 1 - \fft{m}{r} + \fft{c_0(\mu_0 r +1)}{r} e^{-\mu_0 r} - \fft{c_2(\mu_2 r+1)}{2r} e^{-\mu_2 r} + \cdots
\,.
\eea
Here the coefficients $(m,c_0, c_2)$ are coefficients associated with the graviton, massive scalar and massive spin-2 modes.  The masses $(\mu_0, \mu_2)$ of the scalar and spin-2 modes are
\be
\mu_0^2 = \fft{1}{2(3\alpha + \beta)}\,,\qquad \mu_2^2 = -\fft{1}{\beta}\,.
\ee
Thus we see that the excitations of either the massive scalar or the massive spin-2 modes have the effect of yielding $h\ne f$.  Indeed new black holes with $h\ne f$ was constructed numerically in \cite{Lu:2015cqa,Lu:2015psa}. The absence of these two types of modes leads to the Schwarzschild or Schwarzschild-like solutions with $h=f$.  Note that the decoupling of the scalar mode requires that $3\alpha+\beta=0$, corresponding effectively to the Weyl tensor squared in four dimensions.  The decoupling of the massive spin-2 modes requires that $\beta=0$, giving rise to an $f(R)$ theory of gravity.

The cubic ${\cal W}^\3$ (\ref{P31}) we constructed for general dimensions is quasi-topological on any Einstein metrics at the linear level.  In other words, it gives no contribution to the linearized equations of motion.  It follows that the linearized gravity is purely Einstein.  The Schwarzschild black hole, with or without a cosmological constant, is Einstein.  Thus even with the inclusion of the ${\cal W}^\3$ term,  the black hole is ``rigid'' against perturbation provided that we keep the ansatz static with spherical, toroidal or hyperbolic isometries, as in (\ref{genmetric}).  In fact we expect that Schwarzschild black holes are the only solutions from the metric ansatz (\ref{genmetric}) in Ricci gravities we construct in this paper, since none of these theories gives rise to massive scalar or massive spin-2 modes.  However, at the stage of writing this paper, we do not have an opinion whether there exist static black holes outside the class of (\ref{genmetric}) in our quasi-topological Ricci gravities.

\section{Higher-order quasi-topological Ricci gravities}

\subsection{The setup and notations}

In section \ref{sec:cubic}, we studied the properties of the cubic Ricci gravity and obtained quasi-topological combinations.  We now generalize the results to higher-order polynomials. Ricci polynomial invariants are constructed from Ricci scalar $R$ and irreducible Ricci tensor polynomials.  The irreducible Ricci polynomial of $k$'th order is
\be
R_{(k)}=R^{\mu_1}_{\mu_2} R^{\mu_2}_{\mu_3} \cdots R^{\mu_k}_{\mu_1}\,.
\ee
The Ricci scalar can be viewed as $R=R_\1$.  The general Ricci polynomials at the $k$'th order can be expressed as
\be
{\cal R}_{icci}^{(k)} = e_1 R_\1^k + e_2 R^{k-2} R_\2 + e_3 R_\1^{k-3} {R}_\3 +
e_4 R_\1^{k-4} {R}_\2^2 +  e_5 R_\1^{k-4} {R}_\4 + \cdots\,.
\ee
Here we use a lexical ordering of the coupling constants.  The term with higher power of ${R}_{(k)}$ and with smaller $k$ has a smaller labelling index for its coupling constant.  In this paper, we shall work up to the tenth order of Ricci polynomials. The numbers of possible terms for a given order are summarized in Table 1.  They are significantly less than those of the Riemann tensor polynomials at the same order, and the structures are far more manageable.

\bigskip
\begin{table}[ht]
\begin{center}
\begin{tabular}{|c|c|c|c|c|c|c|c|c|c|c|}
  \hline
  $k$ & 1 & 2 & 3 &4 &5 &6 &7 &8 &9 &10  \\ \hline
  $N_k$ & 1 & 2 & 3 &5 & 7 & 11 & 15 & 22 & 30 &42\\
  \hline
\end{tabular}
\caption{\small \it The number $N_{k}$ of possible Ricci polynomial terms at the $k$'th order.}
\end{center}
\end{table}

To derive the equations of motion (\ref{geneom}), a useful formula is
\be
\fft{\partial {R}_{(k+1)}}{\partial R^{\mu\nu\rho\sigma}}=
\ft14 (k+1) \left(g_{\mu\rho} {R}^{(k)}_{\nu\sigma} - g_{\mu\sigma} {R}^{(k)}_{\nu\rho} - g_{\nu\rho} {R}^{(k)}_{\mu\sigma} + g_{\nu\sigma} {R}^{(k)}_{\mu\rho}\right)\,,
\ee
where
\bea
{R}^{\sst{(0)}}_{\mu\nu} &=& g_{\mu\nu}\,,\qquad {R}^\1_{\mu\nu} = R_{\mu\nu}\,,\qquad
{R}_{\mu\nu}^{(2)} = R^{\mu_1}_\mu R_{\nu \mu_1}\,,\nn\\
{R}^{(k)}_{\mu\nu} &=& R^{\mu_1}_\mu R^{\mu_2}_\nu R_{\mu_2}^{\mu_3}\cdots R^{\mu_{k-1}}_{\mu_{k-2}} R_{\mu_{k-1}\mu_1}\,,\qquad k\ge 3\,.
\eea

Since (\ref{genmetric}) is the most general static ansatz with spherical, toroidal or hyperbolic isometries, we can substitute the ansatz directly into the action and obtain the two equations of motion by varying the $(h,f)$ functions. The non-vanishing Ricci tensor components are
\bea
R_{t}^{t} &=& \fft{f}{r} \Big((1-\ft12D) \fft{h'}{h} - \fft{r f' h'}{4fh} +
\fft{r h'^2}{4h^2} - \fft{r h''}{2h} \Big)\,,\nn\\
R_{r}^{r} &=& \fft{f}{r} \Big( (1-\ft12D) \fft{f'}{f} - \fft{r f' h'}{4f h} +
\fft{r h'^2}{4h^2} - \fft{r h''}{2h}\Big)\,,\nn\\
R_{j}^i &=& \Theta\, \delta_j^i \equiv\fft{f}{r} \Big( \fft{(D-3)(\epsilon - f)}{rf} - \fft{f'}{2f} - \fft{h'}{2h}\Big)\delta^i_j\,.
\eea
Thus we have
\be
{R}_{(k)} = (R_{t}^t)^k + (R_{r}^r)^k + (D-2) \Theta^k\,.
\ee
In extended gravities of curvature polynomials (without their covariant derivatives), there are at most four derivatives in equations of motion. By requiring that the equations do not involve higher derivatives, we obtain effective two-derivative theories.  It turns out that this can be achieved, and further more, it has a consequence that the higher-order terms do not contribute to the equations of motion at all, leading to quasi-topological gravities. In the next two subsections, we construct quasi-topological gravities on both the special static metrics (\ref{spemetric}) and the general static metrics (\ref{genmetric}). It turns out that in these theories, the Ricci polynomials all vanish identically with the respective static metric ans\"atze.  This implies that if ${\cal P}^{(k)}$ is quasi-topological, so is $X\, {\cal P}^{(k)}$, where $X$ is any Ricci tensor invariants.  We call these reducible solutions, and we shall present only the irreducible solutions that cannot be factorized.  The reducible solutions can be easily constructed from the irreducible ones, by multiplying an arbitrary polynomial of some desired order. We then construct linearized quasi-topological gravities on Einstein metrics.  In the last subsection, we include Riemann tensor invariants.

\subsection{Quasi-topological gravities for special static metrics}

\label{sec:qtgssm}

For the special static metric ansatz (\ref{spemetric}) in general $D\ne 4$ dimensions, the first occurrence of quasi-topological Ricci gravities is at the order of 4.  The corresponding quartic Ricci polynomial is given by
\bea
{\cal P}^\4 &=& R_\1^4-2 D R_\1^2 R_\2 + 8 (D-2) R_\1 R_\3+\left(D^2-6 D+12\right) R_\2^2\cr
&&-2 (D-2) D R_\4\,.
\eea
(As we have disclaimed in the introduction, it is understood that by ``general dimensions, we mean $D\ge 3$.)  For $k=5$, there are two linearly independent solutions. As we remarked earlier, one is simply the reducible $R_\1 {\cal P}^\4$ and the irreducible solution (that cannot be factorized) is given by
\bea
{\cal P}^\5 &=& R_\1^3 R_\2-D R_\1^2 R_\3-D R_\1 R_\2^2+6 (D-2) R_\1 R_\4\cr
&&+\left(D^2-4 D+8\right) R_\2 R_\3-2 (D-2) D R_\5\,.
\eea
For $k=6$, there are a total five independent solutions, three of which are reducible, namely
\be
(x_1 R_\1^2 + x_2 R_\2) {\cal P}^\4 + x_3 R_\1 {\cal P}^\5\,,\label{preducible}
\ee
for arbitrary constants $x_i$.  The remaining two are irreducible, given by
\bea
{\cal P}^\6_1 &=& R_\1^2 R_\2^2 -2 D R_\1 R_\2 R_\3 +\left(D^2-4 D+8\right) R_\3^2\cr
&& + 2(D-2)\left( 2 R_\1 R_\5 + R_\2 R_\4 -D R_\6\right)\,,\nn\\
{\cal P}^\6_2 &=& R_\1^2 R_\4 -2 R_\1 R_\2 R_\3 + R_\2^3 -D R_\2 R_\4 +D R_\3^2\,.
\eea
We construct irreducible solutions up to the tenth order, and the results are summarized in (\ref{pres}) in appendix \ref{sec:list}.  Note that the reducible solutions associated with coefficients $(x_1,x_2,x_3)$ in (\ref{preducible}) are linearly independent; however, such analogous reducible construction in higher orders may not always be, and hence could potentially lead to a wrong and over counting of total solutions.

The situation in four dimensions is somewhat different.  The lowest occurrence is at the third order, given by (\ref{Q31}).  At each order upward, up to the tenth order, there is a new irreducible solution emerging.  They can be given by the ${\cal P}^{(k)}$ specialized with $D=4$.  However, with the existence of ${\cal Q}^\3$ (\ref{Q31}), we can use it to simplify the higher-order terms further by subtracting some appropriate reducible combinations.  The results are called ${\cal Q}^{(k)}$, and we present them in (\ref{qres}) of appendix \ref{sec:list}.  Note that ${\cal P}^\6_2$ is irreducible in general dimensions, but becomes reducible in $D=4$, owing to the existence of ${\cal Q}^\3$.

It should be emphasized that the reducible quasi-topological Ricci polynomials are no less trivial than the irreducible ones.  In fact they can produce non-trivial perturbations on the background metrics as much as the irreducible solutions.  For a given order, the complete structure includes all the irreducible polynomials as well as the reducible ones.  However for avoiding the proliferation of the results in presentation, we only list the irreducible configurations in appendix \ref{sec:list}.  The full lists of solutions can be easily derived from the irreducible subset. However one needs to be careful to mod out those linearly dependent reducible combinations in order to get a correct counting of all valid linearly independent quasi-topological Ricci polynomials.

    The coefficients of the irreducible polynomial solutions in general dimensions we have
constructed so far are dependent on dimension parameter $D$.  This means that the theories we constructed above are all different in different dimensions.  It would be of interest to construct a theory that is valid in all dimensions.  Such a structure does exist and can be found by selecting appropriate combinations of irreducible and reducible quasi-topological polynomials.  At the eighth order, we find first such an example, given by
\bea
\widehat {\cal P}^\8 &=& R_\2^4-2 R_\1 R_\3 R_\2^2-8 R_\4 R_\2^2+6 R_\3^2 R_\2+8 R_\1 R_\5 R_\2+4 R_\6 R_\2\cr
&&+R_\1^2 R_\3^2+12 R_\4^2-4 R_\1 R_\3 R_\4-16 R_\3 R_\5-2 R_\1^2 R_\6\,.\label{hp-8}
\eea
It is easy to verify that $\widehat {\cal P}^\8$ vanishes identically for the metric ansatz (\ref{spemetric}) in all dimensions.  New irreducible such examples at the ninth and tenth orders are presented in (\ref{hpres}) in appendix \ref{sec:list}.

\subsection{Quasi-topological gravities for general static metrics}

\label{sec:qtggsm}

In this subsection, we construct quasi-topological Ricci polynomials on the general static ansatz (\ref{genmetric}).  The metrics are more general and hence the conditions are more restrictive.  For general $D\ne 4$ dimensions, the first example occurs at the sextic order, given by
\bea
{\cal U}^\6 &=& R_\1^6-3 D R_\1^4 R_\2 + 4 (3 D-5) R_\1^3 R_\3 + 3 \left(D^2-3 D+5\right) R_\1^2 R_\2^2\cr
&&-3 (D-2) (D+5) R_\1^2 R_\4-12 \left(D^2-4 D+5\right) R_\1 R_\2 R_\3\cr
&&+12 (D-2) (D-1) R_\1 R_\5-(D-2) \left(D^2-7 D+15\right) R_\2^3\cr
&&+3 (D-2) \left(D^2-5 D+10\right) R_\2 R_\4+2 \left(3 D^2-15 D+20\right) R_\3^2\cr
&&-2 (D-2) (D-1) D R_\6\,.
\eea
That is to say, the Ricci polynomial ${\cal U}^\6$ gives no contribution to the equations of motion associated with the general static metric ansatz (\ref{genmetric}).  In fact an even stronger statement can be made: substituting (\ref{genmetric}) into ${\cal U}^\6$, it vanishes identically.  We present the full list of irreducible such polynomials in (\ref{ures}) in appendix \ref{sec:list}, up to the tenth order.

In four dimensions, such a polynomial occurs at the quintic order:
\be
{\cal V}^\5 =R_\1^5-10 R_\2 R_\1^3+20 R_\3 R_\1^2+15 R_\2^2 R_\1-30 R_\4 R_\1-20 R_\2 R_\3+24 R_\5\,.
\ee
It is one less than the sextic order for general dimensions, as in the case of quasi-topological gravities on special static metrics (\ref{spemetric}). This allows to simplify the higher-order results further by subtracting appropriate ${\cal V}^\5$ factors.  The irreducible results are given in (\ref{vres}) of appendix \ref{sec:list}.

All the results are dimension dependent, and up to the tenth order, we find no example of $\widehat {\cal U}$  that is independent of dimensions in this rather restrictive case.

\subsection{Linearized quasi-topological gravities on Einstein metrics}

\label{sec:lin}

In this subsection, we consider linearized gravity on Einstein metrics.  It can be a tedious and involved task to perform the linearization of Ricci polynomial gravity on a generic Einstein metric,  as we did in section \ref{sec:cubic} for the cubic invariants.  The upshot in the study of section \ref{sec:cubic} is that there are a total of only two constraints for the theory to be free from the massive scalar and spin-2 modes.  Furthermore, the two constraints for linearized gravity on general Einstein metrics and on maximally symmetric vacua are the same.  We can thus consider the special static ansatz (\ref{spemetric}) and perform linearization $f=g^2 r^2 + \epsilon + \tilde f$.  The absence of higher-derivative terms in the linearized equations of $\tilde f$ yields precise the two constraints.  When these two conditions are satisfied, it turns out that the resulting Ricci polynomials give no contribution to the linearized equations of motion.  The lowest order occurs at $k=3$, and the solution is ${\cal W}^\3$ given in (\ref{P31}).  At the quartic order, there are a total of three solutions: one is reducible $R_\1 {\cal W}^\3$, and the two irreducible ones are
\be
{\cal W}^\4_1 = D R_\4+3 R_\2^2-4 R_\1 R_\3\,,
\qquad
{\cal W}^\4_2 = D^2 R_\4-5 D R_\2^2+4 R_\1^2 R_\2\,.
\ee
For the higher $k$'th order, there are a total of $(N_k - 2)$ solutions, but only one new irreducible solution at each order.  The irreducible solutions, up to the tenth order, are presented in (\ref{wres}) in appendix \ref{sec:list}.

The irreducible ${\cal W}$ series are dependent on dimensions, and hence they are different theories in different dimensions. Together with reducible results, we can build also the dimension-independent structures, with the first two examples at sextic order, given by
\be
\widehat {\cal W}^\6_1 = 3 R_\3^2-4 R_\2 R_\4+R_\1 R_\5\,,\qquad
\widehat {\cal W}^\6_2=2 R_\2^3-3 R_\1 R_\3 R_\2+R_\1^2 R_\4\,.\label{hw6}
\ee
These two theories generate no scalar nor massive spin-2 linear perturbative modes on any Einstein metrics in general dimensions. Higher-order such irreducible polynomials are listed in (\ref{hwres}) of appendix \ref{sec:list}.

\subsection{Including the Riemann tensor}

\label{sec:riemann}

Although we have focused on Ricci tensor polynomials, our results can be generalized to include Riemann tensor polynomials as well, provided that they form an overall factor.  For example for arbitrary $k_1$'th Riemann tensor polynomials, the following direct products of polynomials
\be
R_{\rm iem}^{(k_1)} \left\{{\cal P}^{(k_2)}, {\cal Q}^{(k_2)}, {\cal U}^{(k_2)}, {\cal V}^{(k_2)},
{\cal W}^{(k_2)}\right\}\,,
\ee
give rise to respectively quasi-topological terms at $(k_1+k_2)$ order.  These new topological terms are inequivalent to the original ones, since the Riemann tensor factor can alter the perturbations of the background solutions.

Finally before finishing this section, we would like to comment of a hierarchical structure of our results.  The linearized quasi-topological gravity contains those quasi-topological gravities on special static metrics, which themselves contain those quasi-topological gravities on general static metrics.  Thus our theories are all ghost free at the linear order of perturbations around maximally-symmetric spacetimes and also  Einstein metrics.  (If the theory depends on $D$, then the ghost-free condition is satisfied in $D$ dimensions.)

\section{Lovelock-like quasi-topological Ricci gravities}

In the previous sections, we constructed three types of quasi-topological gravities in general dimensions.  They are all characterized by the fact that they do not contribute to the equations of motion for some restricted class of metric ans\"atze.  Many results were given in terms of a parameter $D$ and the theory is quasi-topological in $D$ dimensions, and they may suffer from having ghost excitations in dimensions other than $D$. We also presented some dimension independent results, in which case the Ricci polynomials are quasi-topological in all dimensions.

These are very different from Lovelock gravities.  The $k$'th-order Lovelock term vanishes identically in $D\le 2k-1$, and is topological in $D=2k$, but nontrivial as well as ghost free in $D>2k$.  Furthermore, the coefficients of Lovelock combinations, although dependent on $k$, are independent of dimensions.  It is of interest to construct Lovelock-like quasi-topological gravities. The cubic example of \cite{Myers:2010ru} is quasi-topological in $D=6$, and nontrivial in $D\ge 5$, but not 6. However the structure depends on dimensions. The Einsteinian cubic gravity is independent of dimensions \cite{Bueno:2016xff}, but never quasi-topological.  In this section, we construct Lovelock-like quasi-topological Ricci gravities that satisfy
the following properties
\begin{itemize}

\item The structure is independent of dimensions.

\item The linearized theory on Einstein metrics in general dimensions is quasi-topological.  In other words, it is simply a linearized Einstein theory, and hence ghost free, with the Newton's constant unmodified.

\item It becomes also quasi-topological on static metrics in one specific dimension $n$.

\end{itemize}
The Ricci polynomials that satisfy the first two criteria are those discussed in section \ref{sec:lin}.  The minimum order of such polynomials is $k=6$ and there are two solutions, given by (\ref{hw6}). It can be easily verified that the combination
\bea
\widetilde {\cal Q}^\6 &=& 2 \widehat {\cal W}^\6_1 + \widehat {\cal W}^\6_2\cr
&=& 2 R_\2^3-3 R_\1 R_\2 R_\3-8 R_\4 R_\2+6 R_\3^2+R_\1^2 R_\4+2 R_\1 R_\5\,,
\eea
is quasi-topological on the special static metrics (\ref{spemetric}) in $D=4$ and only in $D=4$; it is non-trivial and ghost free in other dimensions.  Note that we used notation ${\cal Q}$ to denote quasi-topological polynomials in $D=4$ on special static metrics.  We here use wide tilde on ${\cal Q}$ to denote those Lovelock-like combinations. Both ${\cal Q}$ and $\widetilde {\cal Q}$ satisfy the third criterium above.  The difference is that $\widetilde {\cal Q}$ satisfies the first two criteria as well, in that the theory is ghost free in all dimensions.  The ${\cal Q}$ terms, on the other hand, can develop ghost excitations in dimensions other than four.  The irreducible results of the higher-order $\widetilde {\cal Q}$ terms are given by (\ref{tqres}) in appendix \ref{sec:list}.

      For $n\ne 4$, the lowest order of Lovelock-type of quasi-topological Ricci gravity is $k=7$,
given by
\bea
\widetilde {\cal P}^\7 &=& R_\1 \left(2 R_\2^3-n R_\4 R_\2+2 n R_\3^2\right)+R_\4 R_\1^3-3 R_\2 R_\3 R_\1^2-n R_\2^2 R_\3\cr
&&-(n-2) \left(4 R_\3 R_\4-6 R_\2 R_\5+2 R_\1 R_\6\right)\,.
\eea
The expression has a parameter $n$, but is independent of dimensions $D$ and the linearized gravity on Einstein metrics in any dimension is simply Einstein and hence ghost free, regardless the value of $n$.  In fact the ghost free condition does not even require that $n$ be integer.

For integer $n$, in $D=n$ dimensions, the polynomial ${\cal P}^\7$ becomes quasi-topological on the special static metrics (\ref{spemetric}), but it is nontrivial in any other dimensions.  Higher-order irreducible solutions are given by (\ref{tpres}) in  appendix \ref{sec:list}.  We now have constructed the $({\cal P}, \widehat {\cal P}, \widetilde {\cal P})$ series, all quasi-topological on the special static metrics (\ref{spemetric}).  The ${\cal P}$ series depends on the dimension parameter $D$, and the theories are quasi-topological in $D$ dimensions, but can have ghost excitations in dimensions other than $D$.  The $\widehat {\cal P}$ series is independent of dimensions and is quasi-topological in all dimensions.  The $\widetilde  {\cal P}$ series depends on a parameter $n$, and it is quasi-topological in $D=n$, but nontrivial and ghost free in all other dimensions.

       Lovelock-like quasi-topological gravity exists also on general static metrics
(\ref{genmetric}).  For $n=4$, the lowest order example is
\bea
\widetilde {\cal V}^\8 &=& R_\4 R_\1^4-3 R_\2 R_\3 R_\1^3-R_\5 R_\1^3+2 R_\2^3 R_\1^2+9 R_\3^2 R_\1^2-3 R_\2 R_\4 R_\1^2\cr
&&-2 R_\6 R_\1^2-3 R_\2^2 R_\3 R_\1-16 R_\3 R_\4 R_\1+9 R_\2 R_\5 R_\1+6 R_\7 R_\1\cr
&&-2 R_\2^4-3 R_\2 R_\3^2+12 R_\2^2 R_\4+10 R_\3 R_\5-16 R_\2 R_\6\,.
\eea
Higher-order irreducible solutions are given by (\ref{tvres}) in appendix \ref{sec:list}.  For $n\ne 4$, the lowest order is $k=9$, and there are two solutions:
\bea
\widetilde {\cal U}_1^\9 &=&
-2 \left(3 n^2+n-3\right) R_\3 R_\4 R_\1^2+4 \left(n^2-2 n+3\right) R_\2 R_\5 R_\1^2\cr
&&+\left(15 n^2-65 n+84\right) R_\4^2 R_\1-\left(3 n^2-35 n+57\right) R_\2^2 R_\4 R_\1\cr
&&-4 \left(n^2-13 n+24\right) R_\3 R_\5 R_\1+\left(n^2-n+3\right) R_\2^3 R_\3\cr
&&-2 (n-2) \left(n^2-6 n+18\right) R_\2^2 R_\5-2 (n-2) \left(2 n^2-9 n+15\right) R_\4 R_\5\cr
&&-2 (n-2) \left(n^2-12 n+15\right) R_\2 R_\7+\left(6 n^3-59 n^2+185 n-192\right) R_\2 R_\3 R_\4
\cr
&&-2 n R_\5 R_\1^4+8 n R_\3^2 R_\1^3-4 n R_\2^2 R_\3 R_\1^2+2 (n-2) (n+3) R_\7 R_\1^2\cr
&&-2 n R_\2^4 R_\1+4 (n-3)^2 R_\2 R_\3^2 R_\1-6 (n-2) (n+3) R_\2 R_\6 R_\1\cr
&&-6 (n-2) (n-1) R_\8 R_\1-4 (n-3)^3 R_\3^3+6 (n-3)^2 (n-2) R_\3 R_\6\cr
&&+3 R_\4 R_\1^5-9 R_\2 R_\3 R_\1^4+6 R_\2^3 R_\1^3\,,\cr
\widetilde {\cal U}_2^\9 &=&
-\left(2 n^2-9 n+12\right) R_\3^3-(3 n-2) R_\3 R_\4 R_\1^2+(2 n-1) R_\2 R_\5 R_\1^2\cr
&&+(n-2) R_\7 R_\1^2+(5 n-3) R_\2 R_\3^2 R_\1+(9 n-16) R_\4^2 R_\1\cr
&&-(3 n-1) R_\2^2 R_\4 R_\1-2 (4 n-7) R_\3 R_\5 R_\1-3 (n-2) R_\2 R_\6 R_\1\cr
&&-(n-1) R_\2^3 R_\3+(n-4) (3 n-7) R_\2 R_\3 R_\4-(n-9) (n-2) R_\2^2 R_\5\cr
&&-2 (n-2) n R_\4 R_\5+3 (n-2) n R_\3 R_\6-(n-2) n R_\2 R_\7-R_\5 R_\1^4\cr
&&+R_\3^2 R_\1^3+3 R_\2 R_\4 R_\1^3-5 R_\2^2 R_\3 R_\1^2+2 R_\2^4 R_\1\,,
\eea
At the tenth order, there are four more irreducible solutions, given by (\ref{tures}) in appendix \ref{sec:list}.

\section{Application in the AdS/CFT correspondence}

\label{sec:AdS/CFT}

In this section, we study the effect of the quasi-topological polynomials on the AdS/CFT correspondence, focusing mainly on the linear holographic transport properties.  As was mentioned in the introduction, the Schwarzschild black hole is a solution of the theories, and the quasi-topological terms all give no contribution to the linear perturbations.  It follows that nontrivial effects on the AdS/CFT correspondence from the quasi-topological terms can only arise when additional matter content is included.  As a concrete example, we consider the four-dimensional theory
\be
{\cal L}_4=\sqrt{-g} \left(R - 2\Lambda_0 + \lambda {\cal Q}^\3 + L^{\rm mat}\right)\,,
\label{d4lag}
\ee
where ${\cal Q}^\3$ is given in (\ref{Q31}) and $L^{\rm mat}$ is the matter contribution to the Lagrangian.

The strategy is that we start with (\ref{d4lag}) by setting first $\lambda=0$ and consider matter content such that the Lagrangian admits solutions of special static metric (\ref{spemetric}), e.g. the RN black hole.  We turn on $\lambda$ and the background remains a solution.  We then perform some appropriate perturbations and study the effect of the ${\cal Q}^\3$ term on the relevant holographic properties.

     In this section, we consider the simplest the RN black hole background to study
the effects of ${\cal Q}^\3$ on the holographic shear viscosity (see, e.g.~ \cite{Policastro:2001yc,KSS,KSS0,Buchel:2003tz,Buchel:2004qq,Benincasa:2006fu,
Landsteiner:2007bd,Cremonini:2011iq,Iqbal:2008by,Cai:2008ph,Cai:2009zv,Brustein:2007jj,
Liu:2015tqa,horndeski1,horndeski2,Liu:2016way,Hartnoll:2016tri,Alberte:2016xja,Liu:2016njg},) and thermoelectric DC currents \cite{Donos:2014cya,Amoretti:2014mma,lat6,gsge,DC4,DCsera,Donos:2017oym,Jiang:2017imk,
Baggioli:2017ojd,kim,Wu:2017exh,Liu:2017kml}, and also butterfly velocity \cite{Shenker:2013pqa,Shenker:2013yza,Roberts:2014isa,Shenker:2014cwa,Maldacena:2015waa,
Sfetsos:1994xa,
Blake:2016wvh,Roberts:2016wdl,Blake:2016sud,Reynolds:2016pmi,Lucas:2016yfl,Huang:2016izp,
Ling:2016ibq,Ling:2016wuy,Alishahiha:2016cjk,Feng:2017wvc,Qaemmaqami:2017bdn,Ling:2017jik}.

\subsection{Holographic shear viscosity}
\label{sec:viscosity}

We first include the Maxwell field $A$ and its field strength $F=dA$, with
\be
L^{\rm mat}=-\ft14 F^2\,.
\ee
As was discussed previously, the theory (\ref{d4lag}) admits the general RN AdS black hole.  For our purpose, we consider the AdS planar black hole, given by
\bea
ds^2 &=& - f dt^2 + \fft{dr^2}{f} + r^2 (dx_1^2 + dx_2^2)\,,\qquad A=\fft{q}{r} dt\,,\nn\\
f &=& g^2 r^2 - \fft{\mu}{r} + \fft{q^2}{4r^2}\,.\label{rnadsplanar}
\eea
The solution contains mass and electric charge parameters $\mu$ and $q$ respectively. The parameter $g=1/\ell$ is the inverse of the AdS radius. It can be easily checked that the temperature and entropy density are given by
\be
T=\fft{f'(r_0)}{4\pi}\,,\qquad s=\ft14r_0^2\,,
\ee
where $f(r_0)=0$. In other words, the quasi-topological term gives no contribution to the entropy of the AdS planar black hole.

To study the holographic shear viscosity, we perform the transverse and traceless (TT) perturbation
\be
dx_1^2 + dx_2^2 \rightarrow dx_1^2 + dx_2^2 + 2 \Psi(t,r) dx_1 dx_2\,.
\ee
When $q=0$, the background metric is Einstein, and the ${\cal Q}^\3$ term gives no contribution to the linearized equation.  It follows that $\Psi(r,t)$ is simply a
massless graviton, in which case, the shear viscosity saturates the standard lower bound of the viscosity/entropy ratio \cite{Policastro:2001yc,KSS,KSS0}:
\be
\fft{\eta}{s}=\fft{1}{4\pi}\,.\label{visbound}
\ee
For non-vanishing $q$, the metric is not Einstein, and the higher-derivative term ${\cal Q}^\3$ is no longer ``topological.'' Linear perturbations involve four derivatives and hence contain an extra massive mode.  We make an ansatz
\be
\Psi(t,r) = \varsigma t + \psi(r)\,,
\ee
where $\varsigma$ is a constant.  The linearized equation becomes
\bea
&& \ft12(4g^2 r^3 - \mu) \psi' + \ft18 (4g^2 r^4-4\mu r + q^2) \psi'' +\fft{3\lambda q^2 }{16r^8}\Big[r^2\left(4 g^2 r^4+q^2-4 \mu  r\right)^2\psi''''\cr
&& + 4 r \left(4 g^2 r^4+q^2-4 \mu  r\right) \left(4 g^2 r^4-3 q^2+8 \mu  r\right)\psi'''\nn\\
&& +2  \left(-80 g^4 r^8+112 g^2 \mu  r^5+21 q^4-108 \mu  q^2 r+112 \mu ^2 r^2\right)\psi ''\nn\\
&& -24 \left(-8 g^2 q^2 r^3+36 g^2 \mu  r^4+7 \mu  q^2-24 \mu ^2 r\right)\psi'\Big]=0\,.\label{shearlinear}
\eea
Note that the linear time dependent $\varsigma$ term drops from the equations of motion completely.  As one may have expected, the fourth-order differential equation can be solved by two second order differential equations, namely
\be
\bar\Box \phi_1=0\,,\qquad (\bar\Box-m(r)^2) \phi_2=0\,,\label{phi1phi2}
\ee
with
\be
m(r)^2 = - \fft{1}{6\lambda q^2}r^4\,.\label{vismass}
\ee
Here $\bar \Box$ is the Laplacian with respect to the background  black hole metric.
The most general solution of $\psi$ to (\ref{shearlinear}) is a linear combination of $\phi_1$ and $\phi_2$.  The stability of the system requires that $\lambda<0$ so that $m(r)^2>0$.
The existence of the massive modes implies that the lower bound of the viscosity/entropy ratio is violated \cite{Hartnoll:2016tri}.  (See also \cite{Alberte:2016xja,Liu:2016njg}.)  However, in this case, the massive function $m(r)$ diverges on the boundary as $r\rightarrow \infty$. Consequently, it can be easily verified that we have $\phi_2\sim
\exp[\pm \sqrt{-1/(24 g^2 q^2\lambda)}\, r^2]$ as $r\rightarrow \infty$. This implies that $\phi_2$ should be truncated out by the boundary condition.

Here, we develop a new method to compute the shear viscosity, which we find particularly convenient when higher-derive curvature terms are involved. We consider the radially conserved current constructed in \cite{Liu:2017kml} for general gravity theories, namely
\bea
{\cal J}^{x_1} &=& \sqrt{-g} J^{r x_1}\,,\label{jx1}
\eea
where
\be
J_{}^{\mu\nu} = 2 \fft{\partial L} {\partial R_{\mu\nu\rho\sigma}}
\nabla_\rho\xi_\sigma + 4 \xi_\rho \nabla_\sigma
\fft{\partial L} {\partial R_{\mu\nu\rho\sigma}} + \xi^\rho A_\rho F^{\mu\nu}\,.\label{J2grav}
\ee
For our particular example, the Killing vector is given by $\xi=\partial_{x_2}$.  For the static background, the current ${\cal J}^{x_1}$ vanishes identically; it gives non-trivial contribution at the linear order once the perturbation $\Psi(t,r)$ is turned on.  We find
\be
16\pi {\cal J}_{\rm lin}^{x_1} = g^2 r^2 f\, \psi' + \fft{6\lambda q^2}{r^4} f\,
\Big(r^2 f \psi''' + 2r (r f' - f) \psi'' - 6 (rf' + 2 f - 2g^2 r^2) \psi' \Big)\,.
\ee
It can then be easily verified that the radial conservation law $\partial_r {\cal J}_{\rm lin}^{x_1}=0$ gives precisely the linear equation (\ref{shearlinear}).  In appendix \ref{sec:rccbst}, we show that this radially conserved current matches with the boundary stress tensor $T^{x_1x_2}$, demonstrating that this is indeed the bulk dual of the holographic shear viscosity.

The horizon boundary conditions for $\psi$ must be that $\Psi(t,r)$ is ingoing, implying that
\be
\psi = \varsigma\, \fft{ \log (r-r_0)} {f'(r_0)} + \cdots\,.
\ee
We can now read off the radially conserved current for the $\phi_1$ of (\ref{phi1phi2}) using the horizon data.  We have
\be
J_{\rm lin}^{x_1} (\phi_1)= g^2 r_0^2 \varsigma\,,\qquad\Rightarrow\qquad
\eta =  \fft{1}{g^2} \fft{\partial}{\partial \varsigma} J_{\rm lin}^{x_1}(\phi_1)=
\fft{1}{16\pi}r_0^2\,.
\ee
Thus the shear viscosity/entropy ratio is exactly given by (\ref{visbound}). The quasi-topological term ${\cal Q}^\3$ does not alter the viscosity/entropy ratio. It is of interest to note that
the radially conserved current for the $\phi_2$ mode of (\ref{phi1phi2}) vanishes identically, which implies that the corresponding shear viscosity would vanish. This is of course an irrelevant result since the mode is inconsistent with the boundary condition.

Finally, we would like to comment that the radially-conserved method we adopt in this subsection is particularly convenient to use for calculating the holographic shear viscosity in theories with higher-order curvature terms, owing to the general radially-conserved current formula obtained in \cite{Liu:2017kml}.

\subsection{Thermoelectric DC conductivity}

In order to obtain finite holographic thermoelectric DC conductivities, it is necessary to provide a momentum relaxation mechanism.  A simple way is to introduce two axion fields \cite{EMDC}.
We have therefore the matter system
\be
L^{\rm mat} = -\ft14 F^2 - \ft12 (\partial \chi_1)^2 - \ft12 (\partial \chi_2)^2\,.
\ee
The theory admits the AdS planar black hole \cite{EMDC}
\bea
ds_4^2 &=& - f dt^2 + \fft{dr^2}{f} + r^2 dx^i dx^i \,,\cr
A &=& \fft{q}{r} dt \,, \qquad \chi_i = \beta\, x_i \,,\qquad {i = 1,2} \,,
\eea
with
\be
f = g^2 r^2 - \ft12 \beta^2 - \fft{\mu}{r} + \fft{q^2}{4 r^2} \,.
\ee
The black hole horizon is located at $r_0$, where $f(r_0) = 0$, and the temperature and the entropy density are
\be
T =\fft{f'(r_0)}{4\pi} \,,\qquad s=\ft14 r_0^2\,.
\ee

A small electric field $E_i$ and thermal gradient
$\nabla_i T$ will generate an electric current $J^i$ and thermal
current $Q^i = T^{ti} - \mu J^i$, where $T^{ab} $ is the boundary stress
tensor, $i$ denotes the spatial boundary directions and $\mu$ is the
chemical potential. The conductivity matrix is defined through
\begin{equation}
\left(
  \begin{array}{c}
    J \\
    Q \\
  \end{array}
\right)
=
\left(
  \begin{array}{cc}
    \sigma  & \alpha T  \\
    \bar \alpha T  & \kappa T \\
  \end{array}
  \right)
  \left(
  \begin{array}{c}
    E \\
    - (\nabla T)/T \\
  \end{array}
\right)           \,,
\end{equation}
where $\sigma$ is the electric conductivity, $\bar \kappa$ is the thermal
conductivity and $\alpha\,,\bar \alpha $ are thermoelectric conductivities.

We follow \cite{Donos:2014cya} and consider the linear perturbation
\bea
ds^2 &=& - f dt^2 + \fft{dr^2}{f} + r^2 dx^i dx^i + 2 \Psi_1 dt dx_1 \,,\cr
A &=& a dt + \Psi_2 dx_1 \,, \quad \chi_1 = \beta x_1 + \Psi_3 \,,
\eea
with
\be
\Psi_1 = - t \zeta f + \psi_1(r) \,, \quad \Psi_2 = (- E + \zeta  a) t + \psi_2(r) \,,\qquad
\Psi_3=\psi_3(r)\,.
\ee
A radially conserved electric current can be defined from the Maxwell equation; it is given by
\be
J = \sqrt g F^{rx_1} =-\psi_1 a' + f \psi_2'\,.
\ee
The holographic heat current can be constructed by using the method proposed in \cite{Liu:2017kml}, namely
\bea
Q &=& \sqrt{-g} \Big(  2 \fft{\partial L}{\partial R_{rx_1\rho\sigma}} \nabla_\rho\xi_\sigma + 4 \xi_\rho \nabla_\sigma ( \fft{\partial L}{\partial R_{rx_1\rho\sigma}} ) + a F^{rx_1} \Big)\nn\\
&=& a \left(\psi_1 a' + f \psi_2' \right)+f  \psi_1'-\psi_1 f' \,.
\eea
Note that here $\xi$ is the timelike Killing vector $\partial_t $.  In appendix \ref{sec:rccbst}, we show that current $Q$ matches the relevant boundary stress tensor exactly for general Ricci cubic gravity, and is therefore the bulk dual of the holographic heat current.

Three linear differential equations of $(\psi_1,\psi_2,\psi_3)$ comprise $J'=0$ and $Q'=0$, together with an equation of $\psi_3$ that does not play an essential r\^ole.  What is somewhat surprising is that the quasi-topological term ${\cal Q}^\3$ plays no r\^ole in the linearized equations in this case.  Since both the heat current and electric current are radially conserved, they can be evaluated on the horizon. The ingoing boundary conditions on the horizon are specified by
\bea
\Psi_1 &\sim &  \psi_{10} - \zeta f (t + \int \fft {dr}{ f})  + {\cal O} (r-r_0)+ \dots \,, \cr
 \Psi_2 &\sim & \psi_{20} + (- E + \zeta a ) ( t + \int \fft {dr}{f} ) + {\cal O} (r-r_0)+ \dots \,,
\eea
where $\psi_{10}$ and $\psi_{20}$ are constants. In fact, $\psi_{20}$ is a pure gauge and $\psi_{10}$ can be determined through the equation of motion
\be
\psi_{10} = -\frac{\text{E} q+\zeta  r^2 f'}{\beta^2}\Big|_{r=r_0} \,.
\ee
The electric and heat currents, evaluated on the horizon, are thus given by
\bea
J &=&  \left(1 + \frac{q^2}{\beta ^2 r_0^2}\right) \,E + \frac{4 \pi  q T}{\beta ^2} \,\zeta  \,, \cr
Q &=& \frac{4 \pi  q T}{\beta ^2} \, E + \frac{16 \pi ^2  r_0^2 T^2}{\beta ^2} \,\zeta  \,.
\eea
The elements of thermoelectric DC conductivity matrix can now be obtained
\bea
\sigma &=& \fft{\partial J}{\partial E} = 1 + \frac{q^2}{\beta ^2 r_0^2} \,, \qquad
\alpha = \fft 1 T \fft{\partial J}{\partial \zeta}= \frac{4 \pi  q}{\beta ^2} \,, \cr
\bar \alpha &=& \fft 1 T \fft{\partial Q }{\partial E}= \frac{4 \pi  q}{\beta ^2} \,,\qquad
\bar \kappa = \fft 1 T \fft{\partial Q}{\partial \zeta } = \frac{16 \pi ^2 r_0^2 T }{\beta ^2} \,.
\label{dcmatrix}
\eea
The upshot in this exercise is that the quasi-topological term ${\cal Q}^\3$ plays no role in linearized gravity associated with holographic thermoelectric DC conductivities and the result is simply the same as that of Einstein-Maxwell-Axion theory alone.

\subsection{Butterfly velocity}
\label{sec:butterfly}

In this subsection, we study the butterfly velocity of AdS planar black holes with the metric
\be
ds_4^2 = - f dt^2 + \fft{dr^2}{f} + g^2 r^2 (dx_1^2 + dx_2^2)\,.\label{schwar}
\ee
We refer to the references listed earlier in this section for motivations, and present only the results and minimum steps.  The term ${\cal Q}^\3$ included in our theory does not have any effect on the construction of the background black hole solution.  Near the horizon $r=r_0$, the function $f$ can be expressed as
\be
f=f_1 (r-r_0) + f_2 (r-r_0)^2 + \cdots\,.\label{ftaylor}
\ee
To study the butterfly effects, it is convenient to introduce the Kruskal coordinates $(u,v)$
\be
u=e^{\kappa (r_* - t)}\,,\qquad v=- e^{\kappa (r_*+t)}\,,\qquad
\hbox{with}\qquad dr_* = \fft{dr}{\sqrt{hf}}\,.
\ee
Here $\kappa=2\pi T=\ft12f_1$ is the surface gravity on the horizon $r=r_0$, which corresponds to $uv=0$.  Near the horizon, we have
\be
u v = (r-r_0) - \fft{f_2}{f_1} (r-r_0)^2 + \cdots\,,\qquad
r-r_0 = uv + \fft{f_2}{f_1}\, (uv)^2 + \cdots\,.
\ee
The metric (\ref{schwar}) can now be expressed as
\be
ds_4^2 = A(u v) du dv + B(uv) d x^i d x^i\,,
\ee
where
\be
A(u v) = \fft{1}{\kappa^2} \fft{f}{uv}\,,\qquad B(uv) = g^2 r^2\,.
\ee
We can Taylor expand the functions $A$ and $B$ on the horizon $uv=0$,
\be
A=A_0 + A_1\, (uv) + A_2\, (uv)^2 + \cdots\,,\qquad
B=B_0 + B_1\, (uv) + B_2\, (uv)^2 + \cdots\,.\label{AB}
\ee
The relation between the coefficients $(A_i,B_i)$ and $f_i$ in (\ref{ftaylor}) can be found in \cite{Feng:2017wvc}. To calculate the butterfly velocity, one considers the metric perturbation
\be
ds^2 = A(uv)\,dudv + B(uv)\, dx^i dx^i - A(uv)\,\delta(u)\,h(\vec x)\,du^2.
\ee
We find that the linear butterfly equation of motion for $h(\vec x)$ is
\be
- \fft{12 \lambda A_1}{A_0 B_0^2} ( \Box - \fft{2B_1}{A_0} ) ( \Box - \fft{2B_1}{A_0} - \fft{A_0^2 B_0}{24 \lambda A_1} ) h(\vec x) \sim E e^{2\pi T\, t_w} \delta(\vec x)  \,.\label{butterflyeom}
\ee
Here $\Box$ is the Laplacian of the flat space $dx^i dx^i$. There are two modes with masses
\be
m_0^2 = \fft{2B_1}{A_0} = g^2 r_0 f_1  \,, \quad \tilde m^2 = \fft{2B_1}{A_0} + \fft{A_0^2 B_0}{24 \lambda A_1} = g^2 r_0 f_1 + \fft{g^2 r_0^2}{12 \lambda f_2} \,.\label{m0mtilde}
\ee
Stability requires that both masses squared are non-negative.  The solution to
(\ref{butterflyeom}) takes the form
\be
h(\vec x) \sim  \fft{\tilde E/(m_0^2 - \tilde m^2)}{\sqrt{|\vec x|}}
\Big(e^{2\pi T (t_w - t_*) - m_0 |\vec x|} - e^{2\pi T(t_w - t_*) - \tilde m |\vec x|}\Big)\,.
\ee
It is of important to note that the absence of the derive of the $\delta$-function source implies that the solution no additional integration constant. Since the temperature of the black hole is $T=f_1/(4\pi)$, it follows that $f_1\ge 0$.  This ensures that $m_0^2\ge 0$. To examine the sign of $f_2$, we note that the quasi-topological term $\lambda {\cal Q}^\3$ does not contribute to the background equations of motion. It follows that the black hole solution (\ref{schwar}) is constructed from the standard Einstein equation
\be
G_{\mu\nu} = T_{\mu\nu}^{\rm mat}\,.
\ee
We can thus read off the properties the matter energy-momentum tensor from the Einstein tensor.
For the back hole metric (\ref{schwar}), the matter energy-momentum tensor in the diagonal vielbein base is given by
\be
T^{ab}=
\left(
  \begin{array}{cccc}
    -\fft{f}{r^2} - \fft{f'}{r} &  &  &  \\
     &  \fft{f}{r^2} + \fft{f'}{r} &  &  \\
     &  & \ft12 f'' + \fft{f'}{r} &  \\
     &  &  & \ft12 f'' + \fft{f'}{r} \\
  \end{array}
\right).
\ee
On the horizon, we have $f(r_0)=0$ and $f'(r_0)=f_1\ge 0$, it follows that the null energy condition requires that $f''(r_0)=2f_2\ge 0$.  Note that for the Schwarzschild AdS planar black hole, we have $f''(r_0)=0$ and the $\tilde m$ mode thus decouples.  For the RN AdS planar black hole (\ref{rnadsplanar}), we have $f''(r_0)=q^2/r_0^2\ge 0$.  In general for non-negative values of $\tilde m^2$, the coupling constant $\lambda$ must satisfy
\be
\lambda>0 \qquad \hbox{or}\qquad  \lambda \le -\fft{r_0}{12f_2 f_1}\,.\label{lambdaregion}
\ee
Note that in the context of holographic shear viscosity discussed earlier, the coupling constant has to be negative for the stable TT modes. When the second inequality is saturated, we have $\tilde m=0$.

The butterfly velocity is defined as
\be
v_B = \fft{2 \pi T}{m} \,.
\ee
Thus the corresponding butterfly velocities associated with the $m_0$ and $\tilde m$ modes are given by
\be
v_0^2 = \fft{f_1}{4 g^2 r_0} \,, \qquad \tilde v^2 = \fft{v_0^2}{1 + \fft{r_0}{12\lambda f_1 f_2}} \,.
\ee
The quantity of $v_0$ is the standard butterfly velocity and can be expressed in terms of thermodynamical variables \cite{Feng:2017wvc}
\be
v_0^2 = \fft12 \fft{TS}{V_{\rm th} P}\,.
\ee
Thus we see that for positive $\lambda$, we have $\tilde v^2\le v_0^2$. For some appropriate negative value of $\lambda$, we can have $\tilde v^2 \rightarrow \infty$, which may potentially violate the conjectured holographic diffusivity bounds \cite{elebound,thebound,Blake:2016wvh,Blake:2016sud}. We shall discuss this in the next subsection.

Additional butterfly velocity in Schwarzschild AdS planar black hole of quadratically-extended Ricci gravity was also obtained in \cite{Alishahiha:2016cjk}.  However there is an important difference.  The mass of the new massive mode in \cite{Alishahiha:2016cjk} is solely determined by the coupling constants of the theory, and if it vanishes, it is zero for all solutions.  In our case, $\tilde m$ in (\ref{m0mtilde}) depends not only the coupling constant $\lambda$, but also the integration constant of the solution, such as mass and charges.  Thus in our case, the vanishing of $\tilde m$ depends on the specific property of a solution, provided that $\lambda$ is negative.  This provides a mechanism of breaking the diffusivity bounds, which we shall discuss in the next subsection.

\subsection{Violation of the holographic diffusivity bounds}

In the previous two subsections, we studied both the holographic thermoelectric DC conductivities and butterfly velocities of AdS planar black holes in Einstein-Maxwell-Axion theory in four dimensions, together with the quasi-topological cubic Ricci term $\lambda {\cal Q}^\3$.  We find that the $\lambda Q^\3$ term has no effect on the thermoelectric DC conductivities and yet can give rise a new mode associated with a new butterfly velocity. This phenomenon makes it particularly simple to reexamine how the higher-derivative terms such as $\lambda Q^\3$ affect the conjectured holographic diffusivity bounds.

The diffusivity bound of the Einstein-Maxwell-Axion theory was discussed in \cite{kim}. It is advantageous to write the background electric potential as
\be
a = \mu (1 - \fft{r_0}{r}  ) \,,
\ee
which vanishes on the horizon at $r=r_0$, and the constant $\mu$ can be interpreted as the chemical potential of the boundary CFT. The background metric profile can now be expressed as
\be
f = g^2 r^2-\ft12 {\beta ^2}-\frac{m_0}{r}+\frac{\mu ^2 r_0^2}{4 r^2} \,.
\ee
One can solve for the mass parameter $m_0$ from $f(r_0)=0$ and consequently, the Hawking temperature is
\be
T =\fft{f'(r_0)}{4\pi}= \frac{12 g^2 r_0^2-2 \beta ^2-\mu ^2}{16 \pi  r_0} \,.
\ee
This allows one to express the horizon radius in terms of thermodynamical variables as a simple solution of a quadratic equation:
\be
r_0 = \frac{4 \pi  T + \sqrt{3 g^2 \left(2 \beta ^2+\mu ^2\right)+16 \pi ^2 T^2}}{6 g^2} \,.
\ee
The entropy and charge densities are (We adopt the same unusual convention in literature.)
\bea
S &=& 4 \pi r^2_0 = \frac{\pi  \left(\sqrt{3 g^2 \left(2 \beta ^2+\mu ^2\right)+16 \pi ^2 T^2}+4 \pi  T\right)^2}{9 g^4} \,,\nn\\
\rho &=& \mu r_0 = \frac{\Phi  \left(\sqrt{3 g^2 \left(2 \beta ^2+\mu ^2\right)+16 \pi ^2 T^2}+4 \pi  T\right)}{6 g^2} \,.
\eea
The compressibility and thermoelectric susceptibility are given by
\bea
\chi &=& \Big(\fft{\partial \rho}{\partial \mu}\Big)_T = \fft{1}{6 g^2}  \Big( 4 \pi T +  \frac{6 g^2 \left(\beta ^2+\mu ^2\right)+16 \pi ^2 T^2}{\sqrt{3 g^2 \left(2 \beta ^2+\mu ^2\right)+16 \pi ^2 T^2}} \big)  \,,  \cr
\zeta &=& \Big(\fft{\partial \rho}{\partial T}\Big)_\mu = \frac{2 \pi \mu }{3 g^2} \Big( 1 + \frac{4 \pi T}{\sqrt{3 g^2 \left(2 \beta ^2+\mu ^2\right)+16 \pi ^2 T^2}} \Big) \,.
\eea
The specific heat for fixed chemical potential and for fixed charge density are given by
\bea
c_\mu &=& T (\fft{\partial S}{\partial T})=  \frac{8 \pi ^2 T \left(\sqrt{3 g^2 \left(2 \beta ^2+\mu ^2\right)+16 \pi ^2 T^2}+4 \pi  T\right)^2}{9 g^4 \sqrt{3 g^2 \left(2 \beta ^2+\mu ^2\right)+16 \pi ^2 T^2}} \,, \cr
c_\rho &=&  \frac{4 \pi ^2 T \left(\sqrt{3 g^2 \left(2 \beta ^2+\mu ^2\right)+16 \pi ^2 T^2}+4 \pi  T\right)^3}{9 g^4 \left(3 g^2 \left(\beta ^2+\mu ^2\right)+2 \pi  T \sqrt{3 g^2 \left(2 \beta ^2+\mu ^2\right)+16 \pi ^2 T^2}+8 \pi ^2 T^2\right)} \,.
\eea
The thermoelectric DC conductivities (\ref{dcmatrix}) can also be expressed as functions of $\mu$ and $T$
\bea
\sigma &=& 1 + \frac{\mu ^2}{\beta ^2}  \,, \qquad
\alpha = \bar \alpha = \frac{4 \pi  \rho }{\beta ^2}  = \frac{2 \pi  \mu  \left(\sqrt{3 g^2 \left(2 \beta ^2+\mu ^2\right)+16 \pi ^2 T^2}+4 \pi  T\right)}{3 g^2 \beta ^2} \,, \cr
\bar \kappa &=&  \frac{4 \pi T S }{\beta^2} = \frac{4 \pi ^2 T \left(\sqrt{3 g^2 \left(2 \beta ^2+\mu ^2\right)+16 \pi ^2 T^2}+4 \pi  T\right)^2}{9 g^4 \beta ^2} \,, \cr
\kappa &=& \bar \kappa - \fft{\alpha^2 T}{\sigma} = \frac{16 \pi ^2 r_0^2 T }{\beta ^2+\mu ^2}  = \frac{4 \pi ^2 T \left(\sqrt{3 g^2 \left(2 \beta ^2+\mu ^2\right)+16 \pi ^2 T^2}+4 \pi  T\right)^2}{9 g^4 \left(\beta ^2+\mu ^2\right)} \,.
\eea
The diffusion constants $D_\pm$ are defined by
\be
D_+ D_- = \fft{\sigma}{\chi} \fft{\kappa}{c_\rho} \equiv c_1 \,, \qquad
D_+ + D_- + \fft{\sigma}{\chi} + \fft{\kappa}{c_\rho}  + \fft{T (\zeta \sigma - \chi \alpha )^2}{c_\rho \chi^2 \sigma} \equiv c_2 \,.
\ee
It follows that
\be
D_\pm = \fft{c_2 \pm \sqrt{ c_2^2 - 4 c_1 }}{2} \,.
\ee
These results were given in \cite{kim}.

Owing the presence of the quasi-topological term $\lambda {\cal Q}^\3$, we have shown in the previous subsection that there are two the butterfly velocities.  They can be written also as functions
of $T$ and $\mu$:
\bea
v_0^2 &=& \frac{6 \pi  T}{\sqrt{3 g^2 \left(2 \beta ^2+\mu ^2\right)+16 \pi ^2 T^2}+4 \pi  T}\,, \cr
\tilde v^2 &=& \frac{v_0^2}{1 + \frac{\left(\sqrt{3 g^2 \left(2 \beta ^2+\mu ^2\right)+16 \pi ^2 T^2}+4 \pi  T \right)^3}{5184 \pi  \lambda  g^6 T \left(\beta ^2+\mu ^2\right)}} \,.
\eea
In the incoherent region
\be
\fft \beta T >> 1 \qquad  \text{and }\qquad \fft\beta \mu >>1 \,,
\ee
the diffusion constants and butterfly velocities are given by
\bea
D_+ &=&\frac{\sqrt{6} g}{\beta }-\frac{4 \pi  T}{\beta ^2}  + \frac{3 g^2 \mu ^2+16 \pi ^2 T^2}{2 \sqrt{6} g \beta ^3}+ \dots \,, \cr
D_- &=&\frac{\sqrt{\frac{3}{2}} g}{\beta } +   \frac{16 \pi ^2 T^2-3 g^2 \mu ^2}{4 \sqrt{6} g \beta ^3}+\dots \,, \cr
v_0^2 &=& \frac{\sqrt{6} \pi  T}{g \beta }-\frac{4 \pi ^2 T^2}{g^2 \beta ^2}+\frac{16 \pi ^3 T^3-3 \pi  g^2 \mu ^2 T}{2 \sqrt{6} g^3 \beta ^3}+ \dots \,, \cr
\tilde v^2 &=& \frac{864 \pi ^2 \lambda  g^2 T^2}{\beta ^2}-\frac{2304 \left(\sqrt{6} \pi ^3 \lambda  g T^3 \left(1 + 54 \lambda  g^4\right)\right)}{\beta ^3}+ \dots \,.
\eea
In the incoherent limit, the constants $D_+$, $D_-$ correspond to the charge diffusion constant $D_c$ and energy diffusion constant $D_e$ respectively. It was conjectured that the charge and energy  diffusion constants have the bounds \cite{Blake:2016wvh,Blake:2016sud}
\be
\fft{2 \pi T D_{c,e}}{v_B^2} \ge {\cal C}_{c,e} \,,
\ee
where ${\cal C}_{c,e}$ are pure constants of order one.  To examine this bound in our case, we define
\bea
C_+ &=& \fft{2 \pi T D_+}{v_0^2} =  2 g^2+\frac{g^2 \mu ^2}{\beta ^2}+\frac{2 \sqrt{\frac{2}{3}} \pi  g \mu ^2 T}{\beta ^3}+ \cdots \,, \cr
C_- &=&   \fft{2 \pi T D_-}{v_0^2} = g^2+\frac{2 \sqrt{\frac{2}{3}} \pi  g T}{\beta }+\frac{8 \pi ^2 T^2}{3 \beta ^2}+\frac{\pi  T \left(16 \pi ^2 T^2-9 g^2 \mu ^2\right)}{3 \sqrt{6} g \beta ^3}+ \cdots \,, \cr
\tilde C_+ &=&  \fft{2 \pi T D_+}{\tilde v^2} =   \frac{\beta }{72 \sqrt{6} \pi  \lambda  g T}+\frac{1}{36 \lambda  g^2}+2 g^2+\frac{g^2 \mu ^2+48 \pi ^2 T^2}{288 \sqrt{6} \pi  \lambda  g^3 \beta  T}+ \cdots \,, \label{allcs}\\
\tilde C_- &=&   \fft{2 \pi T D_-}{\tilde v^2} = \frac{\beta }{144 \sqrt{6} \pi  \lambda  g T}+\frac{1}{54 \lambda  g^2}+g^2+\frac{16 \pi ^2 T^2 \left(432 \lambda  g^4+17\right)-3 g^2 \mu ^2}{1728 \sqrt{6} \pi  \lambda  g^3 \beta  T}+ \cdots \,.\nn
\eea
Obviously, the usual mode associated with the butterfly velocity $v_0$ indeed satisfies the bound, with
\be
C_+ \sim 2 g^2  + \cdots \qquad \text{and}\qquad  C_- \sim g^2 + \cdots\,.
\ee
The situation for the quantities $\tilde C_\pm$ is more complicated, depending on the sign of the coupling constant $\lambda$ in the quasi-topological term $\lambda {\cal Q}^\3$.  For positive $\lambda$, the bounds continue to hold since in this case $\tilde C_\pm > C_\pm$.  However,  as was explained in sections \ref{sec:viscosity} and \ref{sec:butterfly}, the coupling constant $\lambda$ can be negative, and should be negative in the discussion of holographic shear viscosity. The bounds can then be violated for negative $\lambda$, for which case the butterfly velocity $\tilde v^2$ is unbound above, leading to a possible infinitesimal $\tilde C_\pm\rightarrow 0^+$.  Although it appears that $\tilde C_\pm$ in (\ref{allcs}) can be negative for negative $\lambda$.  This will not happen since we require (\ref{lambdaregion}) so that $\tilde v^2$ be non-negative, but possible to be unbound above.  Note also that in our model, the violations of diffusivity bounds occur simultaneously for both the diffusion constants $D_c$ and $D_e$. Violations of charge diffusivity bound was previously obtained
\cite{Baggioli:2016oqk,Baggioli:2016oju,Gouteraux:2016wxj,Baggioli:2016pia}, but not for the energy diffusivity bound.

\section{Conclusions}

There are specific combinations of Remann tensor polynomial invariants that give no contributions to the equations of motion in certain dimensions.  These include the Gauss-Bonnet term in four dimensions and more general $k$'th order Lovelock gravities in $D=2k$ dimensions, and the property holds for any metrics in these dimensions. These terms are a total derivative and measure the topological characteristics of the manifold of the corresponding metric.  A quasi-topological term can be defined as a combination of some curvature polynomial invariants that does not contribute to the equations of motion on some special subclass of metrics that one would like to study.  In this paper, we focused on Einstein gravity extended with Ricci tensor polynomial invariants ${\cal R}_{\rm icci}^{(k)}$, up to the tenth order, i.e.~$k=10$.  We constructed quasi-topological gravities for both the general static metric ansatz (\ref{genmetric}) and the special static metric ansatz (\ref{spemetric}).  We also constructed linearized quasi-topological gravities on general Einstein metrics.  We obtained classes of solutions which we label as the
$({\cal W}, \widehat {\cal W}, {\cal P}, \widehat {\cal P}, \widetilde {\cal P}, {\cal Q},
\widetilde {\cal Q}, {\cal U}, \widetilde {\cal U}, {\cal V}, \widetilde {\cal V})$ series.  The meaning of each symbol is summarised as follows.
\begin{itemize}

\item ${\cal R}_{\rm icci}^{(k)}$: The general Ricci polynomial invariants at the $k$'th order.

\item ${\cal W}^{(k)}$ : Linearized quasi-topological terms on Einstein metrics. The structures generally depend on the dimension parameter $D$.  The term with $D$ parameter is quasi-topological and hence ghost free in $D$ dimensions only.  In dimensions other than $D$, the theory can develop ghost excitations.

\item $\widehat {\cal W}^{(k)}$ : A subclass of ${\cal W}^{(k)}$ that are independent of $D$, and hence they are the same theories and quasi-topological at the linear level in all dimensions.

\item ${\cal P}^{(k)}$: These are quasi-topological for special static metric ansatz (\ref{spemetric}).  The structures in general depend on dimensional parameter $D$.

\item $\widehat {\cal P}^{(k)}$: A subclass of ${\cal P}^{(k)}$ that are independent of $D$.

\item $\widetilde {\cal P}^{(k)}$: The polynomials contain a parameter $n$, but independent of dimensions $D$.  The theories are ghost free at the linear level in all dimensions and become quasi-topological for special static metrics (\ref{spemetric}) in $D=n$ dimensions.

\item ${\cal Q}^{(k)}$: These are quasi-topological gravities for the special static metrics in four dimensions, and can develop linear ghosts in other dimensions.

\item $\widetilde {\cal Q}^{(k)}$: These are subclass of ${\cal Q}^{(k)}$ that are quasi-topological in $D=4$, but nontrivial and ghost free in other dimensions.

\item ${\cal U}^{(k)}$: These are quasi-topological for general static metric ansatz (\ref{genmetric}).  The structures in general depend on dimension parameter $D$.

\item $\widetilde {\cal U}^{(k)}$: The polynomials contain a parameter $n$, but are independent of dimensions $D$.  The theories are ghost free at the linear level in all dimensions and become quasi-topological for general static metrics in $D=n$ dimensions.

\item ${\cal V}^{(k)}$: These are quasi-topological gravities for the general static metrics in four dimensions, and can develop linear ghosts in other dimensions.

\item $\widetilde {\cal V}^{(k)}$: These are subclass of ${\cal V}^{(k)}$ that are quasi-topological in $D=4$, but nontrivial and ghost free in other dimensions.

\end{itemize}
In Table 2, we give the total number of all linearly-independent solutions in each series at a given order.  The number in parenthesis denotes the number of irreducible solutions, whose exact structures are presented in appendix \ref{sec:list}.  It is intriguing to notice that the lowest order of quasi-topological terms in four dimensions is typically one order less than those in other dimensions with the same property.  In other words, it appears that it is easier to construct quasi-topological terms in four dimensions.

\bigskip
\begin{table}[ht]
\begin{center}
\begin{tabular}{|c|c|c|c|c|c|c|c|c|c|}
  \hline
$k$ & 2 & 3 &4 &5 &6 &7 &8 &9 &10  \\ \hline
  ${\cal R}^{(k)}_{\rm icci}$ & 2 & 3 &5 & 7 & 11 & 15 & 22 & 30 &42\\
 \hline
${\cal W}^{(k)}$ & 0 & 1(1) &3(2) &5(1) & 9(1) & 13(1) & 20(1) & 28(1) &40(1)\\
  \hline

$\widehat {\cal W}^{(k)}$ & 0 & 0 & 0 & 0 & 2(2) & 4(2) & 10(3) & 15(3) & 25(4)\\
  \hline

${\cal P}^{(k)}$ & 0 & 0 & 1(1) & 2(1) & 5(2) & 8(1) & 14(1) & 21(1) &  32(1)\\
  \hline

$\widehat {\cal P}^{(k)}$ & 0 & 0 & 0 & 0 & 0 & 0 & 1(1) &3(2) & 7(3)\\
  \hline

$\widetilde {\cal P}^{(k)}$ & 0 & 0 & 0 & 0 & 0 & 1(1) & 4(3) &8(3) & 16(4)\\
  \hline

${\cal Q}^{(k)}$ & 0 & 1(1) & 2(1) & 4(1) & 7(1) & 11(1) & 17(1) & 25(1) & 36(1)\\
  \hline

$\widetilde {\cal Q}^{(k)}$ & 0 & 0 & 0 & 0 & 1(1) & 3(2) & 7(3) & 13(3) & 22(4)\\
  \hline

${\cal U}^{(k)}$ & 0 & 0 & 0 & 0 & 1(1) & 2(1) & 5(2) &9(1) & 16(1)\\
  \hline

$\widetilde {\cal U}^{(k)}$ & 0 & 0 & 0 & 0 & 0 & 0 & 0 &2(2) & 7(5)\\
  \hline

${\cal V}^{(k)}$ & 0 & 0 & 0 & 1(1) & 2(1) & 4(1) & 7(1) &12(1) & 19(1)\\
  \hline

$\widetilde {\cal V}^{(k)}$ & 0 & 0 & 0 & 0 & 0 & 0 & 1(1) & 4(3) & 9(4)\\
  \hline

\end{tabular}
\caption{\small \it The total number of linearly-independent quasi-topological polynomials for a given order in each series.  The number in parenthesis denotes the number of irreducible solutions which are presented in appendix \ref{sec:list}. }
\end{center}
\end{table}

It is perhaps a misnomer to call our series quasi-topological since they vanish identically for the corresponding metric ans\"atze, rather than giving rise to total derivatives, as in a typical case of topological terms.  This implies that the ``quasi-topological number'' is always 0.  This property however makes it much easier to construct reducible solutions, even if we are to include the Riemann tensors also.  The numbers given in Table 2 reflects that we have restricted our attentions to Ricci polynomials only.  As was discussed in section \ref{sec:riemann}, the quasi-topological characteristics are maintained if we multiple any factor of Riemann tensor polynomials, and consequently there are much larger number of quasi-topological gravities than those we listed in Table 2.

     One important property of our quasi-topological polynomials is that they
will not affect the solutions of either the special static metrics (\ref{spemetric}) or the general
static metrics (\ref{genmetric}).  But they can give nontrivial effects on the perturbations.
This allows us to study the higher derivative contributions to the AdS/CFT correspondence using
the previously known static backgrounds.  In this paper, considered the Einstein-Maxwell-Axion theory in four dimensions, together with the quasi-topological term $\lambda {\cal Q}^\3$ of (\ref{Q31}).  The theory admits the RN AdS planar black hole.  We examined how $\lambda {\cal Q}^\3$ affected the shear viscosity, thermoelectric DC conductivities and butterfly velocities associated with the bulk AdS planar black hole.  Interestingly, the cubic polynomial $\lambda {\cal Q}^\3$ has no effect on the thermoelectric DC conductivities, but provides additional massive TT modes in both the linear perturbations associated with the shear viscosity and butterfly velocity.  In both cases, the extra mode has a potential of breaking certain previously established bounds.  (In the case of holographic shear viscosity, the extra mode is inconsistent with the boundary conidtion.) Our results make it particularly simple to examine the holographic diffusivity bounds at the presence of higher-derivative terms, since the only new ingredient from our model is an extra butterfly velocity that may be unbound above.  We find that when the coupling constant $\lambda$ is negative, the diffusivity bounds can be violated owing to the possibility of infinitely large butterfly velocity associated with the extra mode induced by the quasi-topological term.  This phenomenon may be quite universal for our quasi-topological Ricci gravities.

Higher-derivative gravities are rather complicated in general, and can be quickly out of hand when the order gets higher. Constructing quasi-topological gravities may be a compromise between dealing with the most general structures and restricting only to the ghost-free Lovelock terms.  Our preliminary investigation of quasi-topological Ricci polynomials indicates that in spite of being much more manageable than the general Riemann tensor polynomials, they are nevertheless quite rich in structures.

\section*{Acknolwedgement}

We are grateful to Xing-Hui Feng and Hyat Huang for useful discussions. Y-Z.L.~and H.L.~are supported in part by NSFC grants No.~11475024, No.~11175269 and No.~11235003. H-S.L.~is supported in part by NSFC grants No.~11305140, No.~11375153, No.~11475148 and No.~11675144.

\appendix

\section{List of quasi-topological Ricci polynomials}

\label{sec:list}

In this paper, we constructed a variety of quasi-topological Ricci polynomial gravities.  These are
the $({\cal W}, \widehat {\cal W}, {\cal P}, \widehat {\cal P}, \widetilde {\cal P}, {\cal Q},
\widetilde {\cal Q}, {\cal U}, \widetilde {\cal U}, {\cal V}, \widetilde {\cal V})$ series.  In the main body of the paper, only the lowest-order examples were given. In this appendix, we present the full list of irreducible polynomials in each series. First we give the quasi-topological terms for special static metric (\ref{spemetric}) for general $D\ne 4$ dimensions. They are given by
\bea
{\cal P}^\4 &=& R_\1^4-2 D R_\1^2 R_\2 + 8 (D-2) R_\1 R_\3+\left(D^2-6 D+12\right) R_\2^2\cr
&&-2 (D-2) D R_\4\,,\nn\\
{\cal P}^\5 &=&
R_\1^3 R_\2-D R_\1^2 R_\3-D R_\1 R_\2^2+6 (D-2) R_\1 R_\4\cr
&&+\left(D^2-4 D+8\right) R_\2 R_\3-2 (D-2) D R_\5\,,\nn\\
{\cal P}^\6_1 &=& R_\1^2 R_\2^2 -2 D R_\1 R_\2 R_\3 +\left(D^2-4 D+8\right) R_\3^2\cr
&& + 2(D-2)\left( 2 R_\1 R_\5 + R_\2 R_\4 -D R_\6\right)\,,\nn\\
{\cal P}^\6_2 &=& R_\1^2 R_\4 -2 R_\1 R_\2 R_\3 + R_\2^3 -D R_\2 R_\4 +D R_\3^2\,,\nn\\
{\cal P}^\7 &=& -D R_\1^2 R_\5+R_1 R_2^3 + D R_\1 R_\3^2 + 2 (D-2) R_\1 R_6-2 D R_\2^2 R_\3
 \cr
&&+ \left(D^2+6 D-12\right) R_\2 R_\5-6 (D-2) R_\3 R_\4-2 (D-2) D R_\7\,,\nn\\
{\cal P}^\8 &=& R_\2^4-2 D R_\2^2 R_\4 + 8 (D-2) R_\2 R_\6 + \left(D^2-6 D+12\right) R_\4^2 \cr
&&-2 (D-2) D R_\8\,,\nn\\
{\cal P}^\9 &=& R_\1 R_\2^2 R_\4 -D R_\1 R_\4^2 + 2 (D-2) R_\1 R_\8 -D R_\2^2 R_\5\cr
&&+ 4 (D-2) R_\2 R_\7  -4 (D-2) R_\3 R_\6 + D^2 R_\4 R_\5 -2 (D-2) D R_\9\,,\nn\\
{\cal P}^{\smallten} &=& D R_\1^2 R_\8-2 D R_\1 R_\2 R_\7 + R_\2^3 R_\4 -\left(D^2-6 D+12\right) R_\2 R_\8\cr
 &&-D R_\3^2 R_\4 + 2 D^2 R_\3 R_\7 -4 (D-2) R_\4 R_\6 -2 (D-2) D R_{\smallten}\,.
\label{pres}
\eea
In four dimensions, the lowest order of the quasi-topological polynomial is (\ref{Q31}), and we can use it to further simplify the higher-order ones by subtracting appropriate reducible ones, leading to the following irreducible solutions
\bea
{\cal Q}^\3 &=&  R_\1^3-6 R_\2 R_\1+8 R_\3\,,\cr
{\cal Q}^\4 &=& R_\2 R_\1^2-4 R_\3 R_\1-2 R_\2^2+8 R_\4\,,\cr
{\cal Q}^\5 &=& -4 R_\2 R_\3+R_\1 \left(R_\2^2-2 R_\4\right)+8 R_\5\,,\cr
{\cal Q}^\6 &=& R_\2^3-6 R_\4 R_\2+8 R_\6\,,\cr
{\cal Q}^\7 &=& R_\3 R_\2^2-4 R_\5 R_\2-2 R_\3 R_\4+8 R_\7\,,\cr
{\cal Q}^\8 &=& R_\4 R_\2^2-4 R_\6 R_\2-2 R_\4^2+8 R_\8\,,\cr
{\cal Q}^\9 &=& R_\3^3-6 R_\6 R_\3+8 R_\9\,,\cr
{\cal Q}^{\smallten} &=& R_\4 R_\3^2-4 R_\7 R_\3-2 R_\4 R_\6+8 R_{\smallten}\,.\label{qres}
\eea
There are also dimensional independent combinations, with the first example occurring at the octic order.  The irreducible solutions are
\bea
\widehat {\cal P}^\8 &=& R_\2^4-2 R_\1 R_\3 R_\2^2-8 R_\4 R_\2^2+6 R_\3^2 R_\2+8 R_\1 R_\5 R_\2+4 R_\6 R_\2\cr
&&+R_\1^2 R_\3^2+12 R_\4^2-4 R_\1 R_\3 R_\4-16 R_\3 R_\5-2 R_\1^2 R_\6\,,\nn\\
\widehat {\cal P}_\1^\9 &=& R_\3 R_\2^3-R_\1 R_\4 R_\2^2-4 R_\5 R_\2^2-R_\1 R_\3^2 R_\2+6 R_\1 R_\6 R_\2
+4 R_\7 R_\2\cr
&&+2 R_\3^3-2 R_\1 R_\4^2+R_\1^2 R_\3 R_\4+8 R_\4 R_\5-12 R_\3 R_\6-2 R_\1^2 R_\7\,,\nn\\
\widehat {\cal P}_\2^\9 &=&R_\3^3-2 R_\2 R_\4 R_\3-R_\1 R_\5 R_\3+R_\1 R_\4^2+R_\2^2 R_\5\,,\nn\\
\widehat {\cal P}_1^{\smallten} &=& R_\6 R_\2^2-R_\4^2 R_\2-R_\3 R_\5 R_\2+R_\3^2 R_\4+R_\1 R_\4 R_\5-R_\1 R_\3 R_\6\,,\nn\\
\widehat {\cal P}_2^{\smallten} &=& -R_\4 R_\2^3+R_\3^2 R_\2^2+8 R_\6 R_\2^2+R_\1 R_\3 R_\4 R_\2-14 R_\3 R_\5 R_\2\cr
&&-2 R_\1 R_\7 R_\2-R_\1 R_\3^3+12 R_\5^2+6 R_\3^2 R_\4+2 R_\1 R_\4 R_\5\cr
&&-16 R_\4 R_\6+4 R_\3 R_\7\,,\nn\\
\widehat {\cal P}_3^{\smallten} &=& -2 R_\4 R_\2^3+3 R_\3^2 R_\2^2+18 R_\6 R_\2^2-2 R_\4^2 R_\2-36 R_\3 R_\5 R_\2+4 R_\8 R_\2\cr
&&-2 R_\1 R_\3^3+R_\1^2 R_\4^2+32 R_\5^2+18 R_\3^2 R_\4+4 R_\1 R_\4 R_\5\cr
&&-36 R_\4 R_\6-2 R_\1^2 R_\8\,.\label{hpres}
\eea

Now we give the quasi-topological terms for the general static metric (\ref{spemetric}) for general $D\ne 4$ dimensions. They are given by
\bea
{\cal U}^\6 &=& R_\1^6-3 D R_\1^4 R_\2 + 4 (3 D-5) R_\1^3 R_\3 + 3 \left(D^2-3 D+5\right) R_\1^2 R_\2^2\cr
&&-3 (D-2) (D+5) R_\1^2 R_\4-12 \left(D^2-4 D+5\right) R_\1 R_\2 R_\3\cr
&&+12 (D-2) (D-1) R_\1 R_\5-(D-2) \left(D^2-7 D+15\right) R_\2^3\cr
&&+3 (D-2) \left(D^2-5 D+10\right) R_\2 R_\4+2 \left(3 D^2-15 D+20\right) R_\3^2\cr
&&-2 (D-2) (D-1) D R_\6\,,\nn\\
{\cal U}^\7 &=&R_\1^5 R_\2 - D R_\1^4 R_\3 -2 D R_\1^3 R_\2^2 + 2 (3 D-5) R_\1^3 R_\4\cr
&&+2 D^2 R_\1^2 R_\2 R_\3-2 (D-2) (D+5) R_\1^2 R_\5\cr
&&+\left(D^2-3 D+5\right) R_\1 R_\2^3-\left(7 D^2-21 D+20\right) R_\1 R_\2 R_\4\cr
 &&- 4 \left(D^2-4 D+5\right) R_\1 R_\3^2+10 (D-2) (D-1) R_\1 R_\6\cr
&&-\left(D^3-7 D^2+21 D-20\right) R_\2^2 R_\3 + 2 (D-2) \left(D^2-4 D+9\right) R_\2 R_\5\cr
&&+ \left(D^3-D^2-10 D+20\right) R_3 R_4 -2 (D-2) (D-1) D R_\7\,,\nn\\
{\cal U}^\8_1 &=&-R_\1^4 R_\4+2 R_\1^3 R_\2 R_\3+2 R_\1^3 R_\5-R_\1^2 R_\2^3+(2 D-5) R_\1^2 R_\2 R_\4\cr
&&-(D+2) R_\1^2 R_\3^2-(D-2) R_\1^2 R_\6-2 (D-4) R_\1 R_\2^2 R_\3-6 R_\1 R_\2 R_\5\cr
&&+2 (D+1) R_\1 R_\3 R_\4+(D-3) R_\2^4-(D-3) (D+2) R_\2^2 R_\4\cr
&&+\left(D^2-2 D-4\right) R_\2 R_\3^2+(D-2) D R_\2 R_\6\cr
&&-2 (D-3) D R_\3 R_\5+(D-4) D R_\4^2\,,\nn\\
{\cal U}^\8_2 &=& R_\1^4 R_\2^2-D R_\1^4 R_\4 + 4 (D-1) R_\1^3 R_\5-2 D R_\1^2 R_\2^3\cr
&&+ 2 \left(D^2+D-5\right) R_\1^2 R_\2 R_\4 -4 (D-1) R_\1^2 R_\3^2\cr
&&-2 (D-2) (D+3) R_\1^2 R_\6+ 4 (D+1) R_\1 R_\2^2 R_\3\cr
&&-4 (D-1) (D+1) R_\1 R_\2 R_\5-4 \left(D^2-6 D+7\right) R_\1 R_\3 R_\4\cr
&&+8 (D-2) (D-1) R_\1 R_\7+\left(D^2-3 D+1\right) R_\2^4\cr
&&-\left(D^3-8 D+4\right) R_\2^2 R_\4+ 4 (D-4) (D-1) R_\2 R_\3^2\cr
&&+2 (D-2) \left(D^2-D+4\right) R_\2 R_\6-4 (D-4) (D-1) R_\3 R_\5\cr
&&+\left(D^3-D^2-14 D+20\right) R_\4^2-2 (D-2) (D-1) D R_\8\,,\nn\\
{\cal U}^\9 &=&-3 D R_\1^4 R_\5+R_\1^3 R_\2^3+3 D R_\1^3 R_\3^2,(9 D-1) R_\1^3 R_\6
\cr
&&+3 \left(2 D^2-2 D-3\right) R_\1^2 R_\2 R_\5-3 (7 D-1) R_\1^2 R_\3 R_\4\cr
&&-3 (D-2) (2 D+1) R_\1^2 R_\7-3 D R_\1 R_\2^4+3 (5 D-2) R_\1 R_\2^2 R_\4\cr
&&-3 \left(3 D^2-7 D-2\right) R_\1 R_\2 R_\3^2-3 \left(D^2+9 D-5\right) R_\1 R_\2 R_\6\cr
&&+6 \left(D^2+D-3\right) R_\1 R_\3 R_\5+9 (2 D-1) R_\1 R_\4^2+6 (D-2) (D-1) R_\1 R_\8\cr
&&+\left(6 D^2-15 D+2\right) R_\2^3 R_\3-3 \left(D^3+D^2-9 D+1\right) R_\2^2 R_\5\cr
&&+3 \left(D^2-3 D-1\right) R_\2 R_\3 R_\4+3 (D-2) D (2 D+1) R_\2 R_\7\cr
&&+ \left(2 D^3-3 D^2-14 D+12\right) R_\3^3-\left(6 D^3-12 D^2-33 D+32\right) R_\3 R_\6\cr
&&+3 (D-1) \left(D^2-3 D-8\right) R_\4 R_\5,-2 (D-2) (D-1) D R_\9\,,\nn\\
{\cal U}^{\smallten} &=& R_\1^2 R_\2^4-4 R_\1^2 R_\2 R_\6-\left(D^2-3 D-1\right) R_\1^2 R_\4^2+ (D-2) (D-1) R_\1^2 R_\8\cr
&&-4 D R_\1 R_\2^3 R_\3 + 12 (D-1) R_\1 R_\2^2 R_\5 + 4 (D-2) (D-1) R_\1 R_\2 R_\3 R_\4\cr
&&-4 (D-2) (D+2) R_\1 R_\2 R_\7-4 \left(D^2-4 D+2\right) R_\1 R_\3 R_\6\cr
&&-12 R_\1 R_\4 R_\5+4 (D-2) (D-1) R_\1 R_\9+D R_\2^5\cr
&&-2 \left(D^2+1\right) R_\2^3 R_\4,2 \left(D^2+2\right) R_\2^2 R_\3^2+2 \left(2 D^2-5 D+4\right) R_\2^2 R_\6\cr
&&-12 \left(D^2-2 D+2\right) R_\2 R_\3 R_\5+\left(D^3-3 D^2+11 D-6\right) R_\2 R_\4^2\cr
&&-(D-6) (D-2) (D-1) R_\2 R_\8-2 (D-2) \left(D^2-4 D+6\right) R_\3^2 R_\4\cr
&&+4 (D-2) \left(D^2-2 D+4\right) R_\3 R_\7-2 \left(3 D^2-6 D+4\right) R_\4 R_\6\cr
&&+6 \left(D^2-3 D+4\right) R_\5^2-2 (D-2) (D-1) D R_{\smallten}\,.\label{ures}
\eea
In this case, there are no combinations whose coefficients are independent of the dimensions.  In other words, up to the tenth order, we find now example of the $\widehat {\cal U}$ series.
In four dimensions, the first example of such polynomial occurs at the quintic order, and it follows we can use it to simplify further the higher-order ${\cal U}^{(k)}$ with $D=4$.  The irreducible results are given by
\bea
{\cal V}^\5 &=&R_\1^5-10 R_\2 R_\1^3+20 R_\3 R_\1^2+15 R_\2^2 R_\1-30 R_\4 R_\1-20 R_\2 R_\3+24 R_\5\,,\cr
{\cal V}^\6 &=&
R_\2 R_\1^4-4 R_\3 R_\1^3-6 R_\2^2 R_\1^2+12 R_\4 R_\1^2+20 R_\3 R_\2 R_\1\cr
&&-24 R_\5 R_\1+3 R_\2^3-8 R_\3^2-18 R_\2 R_\4+24 R_\6\,,\cr
{\cal V}^\7 &=&R_\2^2 R_\1^3-R_4 R_\1^3-6 R_\2 R_\3 R_\1^2+6 R_\5 R_\1^2-3 R_\2^3 R_\1
\cr
&&+6 R_\3^2 R_\1+15 R_\2 R_\4 R_\1-18 R_\6 R_\1+8 R_\2^2 R_\3\cr
&&-14 R_\3 R_\4-18 R_\2 R_\5+24 R_\7\,,\cr
{\cal V}^\8 &=& -R_\2^4+R_\1^2 R_\2^3-6 R_\1 R_\3 R_\2^2+9 R_\4 R_\2^2+6 R_\3^2 R_\2\cr
&&-3 R_\1^2 R_\4 R_\2+12 R_\1 R_\5 R_\2-20 R_\6 R_\2-6 R_\4^2\cr
&&+6 R_\1 R_\3 R_\4-12 R_\3 R_\5+2 R_\1^2 R_\6-12 R_\1 R_\7+24 R_\8\,,\cr
{\cal V}^\9 &=& R_\1 R_\2^4-6 R_\1 R_\2^2 R_\4+8 R_\1 R_\2 R_\6+3 R_\1 R_\4^2-6 R_\1 R_\8-4 R_\2^3 R_\3\cr
 &&+ 12 R_\2^2 R_\5 +12 R_\2 R_\3 R_\4-24 R_\2 R_\7-8 R_\3 R_\6-12 R_\4 R_\5+24 R_\9\,,\cr
{\cal V}^{\smallten} &=& R_\2^5-10 R_\2^3 R_\4+20 R_\2^2 R_\6 + 15 R_\2 R_\4^2-30 R_\2 R_\8\cr
&&-20 R_\4 R_\6 + 24 R_{\smallten}\,.\label{vres}
\eea

We now present the linearized quasi-topological gravities on Einstein metrics.  Up to the tenth order, the irreducible solutions are given by
\bea
{\cal W}^\3 &=& 2D^2 R_\3-3D R_\2 R_\1+2R_\1^3\,,\qquad
{\cal W}^\4_1 = D R_\4+3 R_\2^2-4 R_\1 R_\3\,,\cr
\qquad
{\cal W}^\4_2 &=& D^2 R_\4-5 D R_\2^2+4 R_\1^2 R_\2\,,\qquad
{\cal W}^\5 = D R_\5+2 R_\2 R_\3-3 R_\1 R_\4\,,\cr
{\cal W}^\6 &=& 4 D R_\6+5 R_\3^2-9 R_\1 R_\5\,,\qquad
{\cal W}^\7 = D R_\7+R_\3 R_\4-2 R_\1 R_\6\,,\cr
{\cal W}^\8 &=& 12 D R_\8+7 R_\4^2-16 R_\1 R_\7\,,\quad
{\cal W}^\9 = 35 D R_\9+18 R_\4 R_\5-45 R_\1 R_\8\,,\cr
{\cal W}^{\smallten} &=& 20 D R_{\smallten}+9 R_\5^2-25 R_\1 R_\9\,.\label{wres}
\eea
The structures depend on the parameter $D$ and the linearized gravity on $D$-dimensional Einstein metrics for each term simply vanishes.  Linear ghost modes can emerge in dimensions other than $D$.
Together with the reducible solutions, we con construct linearized quasi-topological gravities
that are independent of dimensions. The irreducible dimension-independent combinations are given by
\bea
\widehat {\cal W}^\6_1 &=& 3 R_\3^2-4 R_\2 R_\4+R_\1 R_\5\,,\qquad
\widehat {\cal W}^\6_2=2 R_\2^3-3 R_\1 R_\3 R_\2+R_\1^2 R_\4\,,\cr
\widehat {\cal W}^\7_1 &=& 2 R_\3 R_\4-3 R_\2 R_\5+R_\1 R_\6\,,\qquad
\widehat {\cal W}^\7_2 = R_\3 R_\2^2+R_\1 R_\4 R_\2-2 R_\1 R_\3^2\,,\cr
\widehat {\cal W}^\8_1 &=& 3 R_\4^2-4 R_\3 R_\5+R_\2 R_\6\,,\qquad
\widehat {\cal W}^\8_2 = 8 R_\4^2-9 R_\3 R_\5+R_\1 R_\7\,,\cr
\widehat {\cal W}^\8_3 &=& -2 R_\4 R_\2^2+R_\3^2 R_\2+R_\1 R_\3 R_\4\,,\qquad
\widehat {\cal W}^\9_1 = 2 R_\4 R_\5-3 R_\3 R_\6+R_\2 R_\7\,,\cr
\widehat {\cal W}^\9_2 &=& 5 R_\4 R_\5-6 R_\3 R_\6+R_\1 R_\8\,,\qquad
\widehat {\cal W}^\9_3 = 2 R_\3^3-3 R_\2 R_\4 R_\3+R_\2^2 R_\5\,,\cr
\widehat {\cal W}^{\smallten}_1 &=& 15 R_\5^2-16 R_\4 R_\6+R_\1 R_\9\,,\qquad
\widehat {\cal W}^{\smallten}_2 = 3 R_\5^2-4 R_\4 R_\6+R_\3 R_\7\,,\cr
\widehat {\cal W}^{\smallten}_3 &=& 8 R_\5^2-9 R_\4 R_\6+R_\2 R_\8\,,\qquad
\widehat {\cal W}^{\smallten}_4 = R_\6 R_\2^2-5 R_\4^2 R_\2+4 R_\3^2 R_\4\,.\label{hwres}
\eea
Any of the above theories is quasi-topological at the linear level in all dimensions.

Finally we present the irreducible solutions of Lovelock-like quasi-topological solutions.
These theories are independent of dimensions and have one parameter $n$.  The linearized theories on Einstein metrics are simply Einstein and hence ghost free, in all dimensions.  In $D=n$ dimensions, the theories become
quasi-topological.  The irreducible solutions with $n\ne 4$ on special static metrics (\ref{spemetric}) are
\bea
\widetilde {\cal P}^\7 &=& R_\1 \left(2 R_\2^3-n R_\4 R_\2+2 n R_\3^2\right)+R_\4 R_\1^3-3 R_\2 R_\3 R_\1^2-n R_\2^2 R_\3\cr
&&-(n-2) \left(4 R_\3 R_\4-6 R_\2 R_\5+2 R_\1 R_\6\right)\,,\cr
\widetilde {\cal P}^\8_1 &=& 3 n R_\4^2-4 n R_\3 R_\5+R_\2 \left(n R_\6+3 R_\3^2+4 R_\1 R_\5\right)-R_\6 R_\1^2\cr
&&-2 R_\1 R_\3 R_\4-4 R_\2^2 R_\4
\,,\cr
\widetilde {\cal P}^\8_2 &=& -R_\2^2 \left(2 n R_\4+3 R_\1 R_\3\right)+R_\2 \left(n R_\3^2+R_\4 R_\1^2\right)+n R_\1 R_\3 R_\4\cr
&&+(n-2) \left(-4 R_\4^2+2 R_\3 R_\5+4 R_\2 R_\6-2 R_\1 R_\7\right)+2 R_\2^4
\,,\cr
\widetilde {\cal P}^\8_3 &=& \left(n^2+40 n-80\right) R_\6 R_\2+3 \left(n^2-24 n+48\right) R_\4^2-20 n R_\4 R_\2^2\cr
&&-4 \left(n^2-14 n+28\right) R_\3 R_\5+7 n R_\3^2 R_\2+14 n R_\1 R_\3 R_\4-n R_\1^2 R_\6\cr
&&-24 (n-2) R_\1 R_\7+20 R_\2^4-24 R_\1 R_\3 R_\2^2+4 R_\1^3 R_\5\,,
\cr
\widetilde {\cal P}^\9_1 &=& -R_\3 R_\2 \left(3 n R_\4+2 R_\1 R_\3\right)+n \left(2 R_\3^3+R_\1 R_\4^2\right)\cr
&&+(n-2) \left(2 R_\4 R_\5-6 R_\3 R_\6+6 R_\2 R_\7-2 R_\1 R_\8\right)\cr
&&+R_\3 R_\2^3+R_\1 R_\4 R_\2^2
\,,\cr
\widetilde {\cal P}^\9_2 &=& n \left(-2 R_\4 R_\5+3 R_\3 R_\6-R_\2 R_\7\right)-R_\3^3+R_\1 R_\4^2+2 R_\2^2 R_\5\cr
&&-3 R_\1 R_\2 R_\6+R_\1^2 R_\7
\,,\cr
\widetilde {\cal P}^\9_3 &=& -n R_\5 R_\2^2+R_\2 \left(-6 n R_\3 R_\4+R_\5 R_\1^2-5 R_\3^2 R_\1\right)\cr
&&+n \left(4 R_\3^3+3 R_\1 R_\4^2\right)+4 R_\3 R_\2^3\cr
&&-2 (n-2) \left(R_\4 R_\5+3 R_\3 R_\6-7 R_\2 R_\7+3 R_\1 R_\8\right)
\,,\cr
\widetilde {\cal P}^\smallten_1 &=& n \left(2 R_\4^2+R_\3 R_\5\right) R_\2-2 n R_\3^2 R_\4+R_\1 \left(R_\3^3-n R_\4 R_\5\right)\cr
&&+(n-2) \left(-2 R_\4 R_\6+6 R_\3 R_\7-6 R_\2 R_\8+2 R_\1 R_\9\right)-R_\2^3 R_\4
\,,\cr
\widetilde {\cal P}^\smallten_2 &=& n \left(-3 R_\5^2+4 R_\4 R_\6-R_\3 R_\7\right)-4 R_\6 R_\2^2+7 R_\3 R_\5 R_\2\cr
&&+R_\1 R_\7 R_\2-3 R_\3^2 R_\4-R_\1 R_\4 R_\5
\,,\cr
\widetilde {\cal P}^\smallten_3 &=& \left(-3 R_\3^2+4 R_\2 R_\4-R_\1 R_\5\right) \left(n R_\4-R_\2^2\right)\cr
&&+2 (n-2) \left(-5 R_\5^2+4 R_\4 R_\6+4 R_\3 R_\7-4 R_\2 R_\8+R_\1 R_\9\right)
\,,\cr
\widetilde {\cal P}^\smallten_4 &=& n \left(-8 R_\5^2+9 R_\4 R_\6-R_\2 R_\8\right)+R_\8 R_\1^2-2 R_\4 R_\5 R_\1\cr
&&+R_\2 R_\4^2-9 R_\3^2 R_\4+18 R_\2 R_\3 R_\5-9 R_\2^2 R_\6
\,.\label{tpres}
\eea
For $n=4$, the irreducible solutions become
\bea
\widetilde {\cal Q}^\6 &=& 2 R_\2^3-3 R_\1 R_\2 R_\3-8 R_\4 R_\2+6 R_\3^2+R_\1^2 R_\4+2 R_\1 R_\5\,,\cr
\widetilde {\cal Q}^\7_1 &=& R_\3 R_\2^2+R_\1 R_\4 R_\2-6 R_\5 R_\2-2 R_\1 R_\3^2+4 R_\3 R_\4+2 R_\1 R_\6\,,\cr
\widetilde {\cal Q}^\7_2 &=& R_\5 R_\1^2-5 R_\3^2 R_\1+6 R_\6 R_\1+4 R_\2^2 R_\3+12 R_\3 R_\4-18 R_\2 R_\5\,,\cr
\widetilde {\cal Q}^\8_1 &=& -2 R_\4 R_\2^2+R_\3^2 R_\2+6 R_\6 R_\2+2 R_\4^2+R_\1 R_\3 R_\4\cr
&&-6 R_\3 R_\5-2 R_\1 R_\7\,,\cr
\widetilde {\cal Q}^\8_2 &=& -4 R_\4 R_\2^2+3 R_\3^2 R_\2+R_\1 R_\5 R_\2+8 R_\6 R_\2+8 R_\4^2\cr
&&-14 R_\3 R_\5-2 R_\1 R_\7 \,,\cr
\widetilde {\cal Q}^\8_3 &=& R_\6 R_\1^2-4 R_\7 R_\1+7 R_\2 R_\3^2+16 R_\4^2-8 R_\2^2 R_\4\cr
&&-28 R_\3 R_\5+16 R_\2 R_\6 \,,\cr
\widetilde {\cal Q}^\9_1 &=& 2 R_\3^3-3 R_\2 R_\4 R_\3-6 R_\6 R_\3+R_\2^2 R_\5+4 R_\4 R_\5+2 R_\2 R_\7 \,,\cr
\widetilde {\cal Q}^\9_2 &=& 2 R_\3^3-3 R_\2 R_\4 R_\3-6 R_\6 R_\3+R_\1 R_\4^2+2 R_\4 R_\5
\cr
&&+6 R_\2 R_\7-2 R_\1 R_\8 \,,\cr
\widetilde {\cal Q}^\9_3 &=& 6 R_\3^3-7 R_\2 R_\4 R_\3-24 R_\6 R_\3+14 R_\4 R_\5+R_\1 R_\2 R_\6\cr
&&+12 R_\2 R_\7-2 R_\1 R_\8\,,\cr
\widetilde {\cal Q}^{\smallten}_1 &=& R_\6 R_\2^2-5 R_\4^2 R_\2+6 R_\8 R_\2+4 R_\3^2 R_\4+10 R_\4 R_\6-16 R_\3 R_\7\,,\cr
\widetilde {\cal Q}^{\smallten}_2 &=& 3 R_\4 R_\3^2-12 R_\7 R_\3-4 R_\2 R_\4^2-2 R_\5^2+R_\1 R_\4 R_\5+8 R_\4 R_\6\cr
&&+8 R_\2 R_\8-2 R_\1 R_\9\,,\cr
\widetilde {\cal Q}^{\smallten}_3 &=& 5 R_\4 R_\3^2+R_\1 R_\6 R_\3-22 R_\7 R_\3-6 R_\2 R_\4^2+12 R_\4 R_\6\cr
&&+12 R_\2 R_\8-2 R_\1 R_\9 \,,\cr
\widetilde {\cal Q}^{\smallten}_4 &=& -R_\4 R_\2^3+12 R_\8 R_\2+R_\1 R_\3^3+4 R_\4 R_\6-12 R_\3 R_\7-4 R_\1 R_\9 \,.\label{tqres}
\eea
Lovelock-like quasi-topological Ricci gravities on general static metric (\ref{genmetric}) with $n\ne 4$ can be constructed from the following irreducible ones:
\bea
\widetilde {\cal U}_1^\9 &=&
-2 \left(3 n^2+n-3\right) R_\3 R_\4 R_\1^2+4 \left(n^2-2 n+3\right) R_\2 R_\5 R_\1^2\cr
&&+\left(15 n^2-65 n+84\right) R_\4^2 R_\1-\left(3 n^2-35 n+57\right) R_\2^2 R_\4 R_\1\cr
&&-4 \left(n^2-13 n+24\right) R_\3 R_\5 R_\1+\left(n^2-n+3\right) R_\2^3 R_\3\cr
&&-2 (n-2) \left(n^2-6 n+18\right) R_\2^2 R_\5-2 (n-2) \left(2 n^2-9 n+15\right) R_\4 R_\5\cr
&&-2 (n-2) \left(n^2-12 n+15\right) R_\2 R_\7+\left(6 n^3-59 n^2+185 n-192\right) R_\2 R_\3 R_\4
\cr
&&-2 n R_\5 R_\1^4+8 n R_\3^2 R_\1^3-4 n R_\2^2 R_\3 R_\1^2+2 (n-2) (n+3) R_\7 R_\1^2\cr
&&-2 n R_\2^4 R_\1+4 (n-3)^2 R_\2 R_\3^2 R_\1-6 (n-2) (n+3) R_\2 R_\6 R_\1\cr
&&-6 (n-2) (n-1) R_\8 R_\1-4 (n-3)^3 R_\3^3+6 (n-3)^2 (n-2) R_\3 R_\6\cr
&&+3 R_\4 R_\1^5-9 R_\2 R_\3 R_\1^4+6 R_\2^3 R_\1^3\,,\cr
\widetilde {\cal U}_2^\9 &=&
-\left(2 n^2-9 n+12\right) R_\3^3-(3 n-2) R_\3 R_\4 R_\1^2+(2 n-1) R_\2 R_\5 R_\1^2\cr
&&+(n-2) R_\7 R_\1^2+(5 n-3) R_\2 R_\3^2 R_\1+(9 n-16) R_\4^2 R_\1\cr
&&-(3 n-1) R_\2^2 R_\4 R_\1-2 (4 n-7) R_\3 R_\5 R_\1-3 (n-2) R_\2 R_\6 R_\1\cr
&&-(n-1) R_\2^3 R_\3+(n-4) (3 n-7) R_\2 R_\3 R_\4-(n-9) (n-2) R_\2^2 R_\5\cr
&&-2 (n-2) n R_\4 R_\5+3 (n-2) n R_\3 R_\6-(n-2) n R_\2 R_\7-R_\5 R_\1^4\cr
&&+R_\3^2 R_\1^3+3 R_\2 R_\4 R_\1^3-5 R_\2^2 R_\3 R_\1^2+2 R_\2^4 R_\1\,,\cr
\widetilde {\cal U}^{\smallten}_1 &=&
(n+4) R_\4 R_\2^3-(n+3) R_\3^2 R_\2^2-2 n R_\6 R_\2^2-2 (n+2) R_\4^2 R_\2\cr
&&-(n-4) R_\1 R_\3 R_\4 R_\2+(5 n+3) R_\3 R_\5 R_\2+(n+1) R_\1 R_\7 R_\2\cr
&&+n R_\1 R_\3^3-3 n R_\5^2-(n-1) R_\3^2 R_\4+(n+3) R_\1 R_\4 R_\5\cr
&&-2 (n+2) R_\1 R_\3 R_\6+4 n R_\4 R_\6-n R_\3 R_\7-R_\2^5+3 R_\1 R_\3 R_\2^3\cr
&&-R_\1^2 R_\4 R_\2^2-7 R_\1 R_\5 R_\2^2-2 R_\1^2 R_\3^2 R_\2+4 R_\1^2 R_\6 R_\2-2 R_\1^2 R_\4^2\cr
&&+R_\1^3 R_\3 R_\4+R_\1^2 R_\3 R_\5-R_\1^3 R_\7\,,\cr
\widetilde {\cal U}^{\smallten}_2 &=&
-2 \left(n^2-n+4\right) R_\4^2 R_\2+\left(n^2+10 n-9\right) R_\3 R_\5 R_\2\cr
&&+\left(n^2-5 n+7\right) R_\3^2 R_\4+2 (2 n+1) R_\4 R_\2^3-(3 n+2) R_\3^2 R_\2^2\cr
&&-(n+8) R_\1 R_\5 R_\2^2-4 (2 n-3) R_\6 R_\2^2-(2 n-9) R_\1 R_\3 R_\4 R_\2\cr
&&+(4 n-5) R_\1 R_\7 R_\2+(n-2) n R_\8 R_\2+(n-2) R_\1 R_\3^3\cr
&&+(2 n-5) R_\1^2 R_\4^2-n (n+1) R_\5^2-(n-3) R_\1^2 R_\3 R_\5\cr
&&-(3 n-11) R_\1 R_\4 R_\5+(n-10) R_\1 R_\3 R_\6+n (3 n-2) R_\4 R_\6\cr
&&-n (3 n-5) R_\3 R_\7-(n-2) R_\1^2 R_\8-2 R_\2^5+5 R_\1 R_\3 R_\2^3\cr
&&-3 R_\1^2 R_\4 R_\2^2-R_\1^2 R_\3^2 R_\2+R_\1^3 R_\5 R_\2+4 R_\1^2 R_\6 R_\2-R_\1^3 R_\7\,,\cr
\widetilde {\cal U}^{\smallten}_3 &=&
\left(n^2-28 n+36\right) R_\6 R_\2^2-\left(5 n^2-8 n+36\right) R_\4^2 R_\2\cr
&&+\left(4 n^2-19 n+24\right) R_\3^2 R_\4+2 (5 n+6) R_\4 R_\2^3-(7 n+11) R_\3^2 R_\2^2\cr
&&-2 (4 n-13) R_\1 R_\3 R_\4 R_\2+18 (2 n-1) R_\3 R_\5 R_\2\cr
&&-(2 n-17) R_\1^2 R_\6 R_\2+12 (n-1) R_\1 R_\7 R_\2+3 (n-2) n R_\8 R_\2\cr
&&+2 (n-1) R_\1 R_\3^3+(5 n-14) R_\1^2 R_\4^2-12 n R_\5^2-6 (2 n-7) R_\1 R_\4 R_\5\cr
&&+6 (n-7) R_\1 R_\3 R_\6+n (5 n+6) R_\4 R_\6-4 n (2 n-3) R_\3 R_\7\cr
&&-3 (n-2) R_\1^2 R_\8-6 R_\2^5+14 R_\1 R_\3 R_\2^3-6 R_\1^2 R_\4 R_\2^2-30 R_\1 R_\5 R_\2^2\cr
&&-3 R_\1^2 R_\3^2 R_\2+6 R_\1^2 R_\3 R_\5+R_\1^4 R_\6-4 R_\1^3 R_\7\,,\cr
\widetilde {\cal U}^{\smallten}_4 &=&
\left(n^2-17 n+22\right) R_\6 R_\2^2+\left(3 n^2-23 n+18\right) R_\4^2 R_\2\cr
&&-\left(4 n^2-44 n+46\right) R_\3 R_\5 R_\2+2 \left(2 n^2-17 n+20\right) R_\5^2\cr
&&-\left(7 n^2-52 n+60\right) R_\4 R_\6+2 \left(2 n^2-11 n+12\right) R_\3 R_\7-n R_\2^5\cr
&&+2 n R_\1 R_\3 R_\2^3+(8 n-4) R_\4 R_\2^3-3 (2 n-1) R_\3^2 R_\2^2-2 (n+5) R_\1 R_\5 R_\2^2\cr
&&-n R_\1^2 R_\3^2 R_\2-2 (4 n-9) R_\1 R_\3 R_\4 R_\2-(n-10) R_\1^2 R_\6 R_\2\cr
&&+2 (5 n-7) R_\1 R_\7 R_\2-(n-2)^2 R_\8 R_\2+2 (3 n-4) R_\1 R_\3^3\cr
&&-3 (n-1) R_\1^2 R_\4^2+(8-5 n) R_\3^2 R_\4+2 (3 n-5) R_\1^2 R_\3 R_\5\cr
&&+2 (4 n-3) R_\1 R_\4 R_\5-16 (n-1) R_\1 R_\3 R_\6-(n-2) R_\1^2 R_\8\cr
&&+R_\1^2 R_\2^4-2 R_\1^3 R_\3 R_\2^2+R_\1^4 R_\3^2-2 R_\1^3 R_\7\,,\cr
\widetilde {\cal U}^{\smallten}_5 &=&
\left(n^2+6 n-6\right) R_\4 R_\2^3-2 \left(n^2+6 n-14\right) R_\6 R_\2^2-2 \left(2 n^2-5 n+8\right) R_\4^2 R_\2\cr
&&-\left(2 n^2-31 n+43\right) R_\3 R_\5 R_\2-\left(n^2-4 n+6\right) R_\1 R_\3^3\cr
&&+\left(2 n^2-17 n+20\right) R_\5^2+\left(5 n^2-23 n+29\right) R_\3^2 R_\4\cr
&&+\left(3 n^2-3 n-14\right) R_\1 R_\3 R_\6+\left(n^2+10 n-16\right) R_\4 R_\6\cr
&&-\left(8 n^2-23 n+16\right) R_\3 R_\7-n R_\2^5-5 (n-1) R_\3^2 R_\2^2-2 n R_\1^2 R_\4 R_\2^2\cr
&&+(6 n-17) R_\1 R_\5 R_\2^2 +3 n R_\1^2 R_\3^2 R_\2-(9 n-19) R_\1 R_\3 R_\4 R_\2\cr
&&-(n-6) R_\1^2 R_\6 R_\2+(n+1) R_\1 R_\7 R_\2+(n-2) (7 n-8) R_\8 R_\2\cr
&&+(3 n-7) R_\1^2 R_\4^2-(4 n-7) R_\1^2 R_\3 R_\5-(5 n-17) R_\1 R_\4 R_\5\cr
&&+(n-2) R_\1^2 R_\8-2 (n-2) (n-1) R_\1 R_\9+2 R_\1^2 R_\2^4-3 R_\1^3 R_\3 R_\2^2\cr
&&+R_\1^4 R_\4 R_\2-R_\1^3 R_\7\,.\label{tures}
\eea
The irreducible solutions for $n=4$ are given by
\bea
\widetilde {\cal V}^\8 &=& R_\4 R_\1^4-3 R_\2 R_\3 R_\1^3-R_\5 R_\1^3+2 R_\2^3 R_\1^2+9 R_\3^2 R_\1^2-3 R_\2 R_\4 R_\1^2\cr
&&-2 R_\6 R_\1^2-3 R_\2^2 R_\3 R_\1-16 R_\3 R_\4 R_\1+9 R_\2 R_\5 R_\1+6 R_\7 R_\1\cr
&&-2 R_\2^4-3 R_\2 R_\3^2+12 R_\2^2 R_\4+10 R_\3 R_\5-16 R_\2 R_\6
\,,\cr
\widetilde {\cal V}^\9_1 &=& R_\1 R_\2^4-2 R_\3 R_\2^3-2 R_\1^2 R_\3 R_\2^2-2 R_\1 R_\4 R_\2^2+4 R_\5 R_\2^2\cr
&&+5 R_\1 R_\3^2 R_\2+4 R_\1^2 R_\5 R_\2-5 R_\1 R_\6 R_\2-2 R_\7 R_\2-2 R_\3^3+R_\1^3 R_\3^2\cr
&&+5 R_\1 R_\4^2-4 R_\1^2 R_\3 R_\4-4 R_\1 R_\3 R_\5-4 R_\4 R_\5-R_\1^3 R_\6\cr
&&+6 R_\3 R_\6+2 R_\1^2 R_\7
\,,\cr
\widetilde {\cal V}^\9_2 &=& 2 R_\1 R_\2^4-5 R_\3 R_\2^3-3 R_\1^2 R_\3 R_\2^2-9 R_\1 R_\4 R_\2^2+15 R_\5 R_\2^2\cr
&&+12 R_\1 R_\3^2 R_\2+R_\1^3 R_\4 R_\2+2 R_\3 R_\4 R_\2+3 R_\1^2 R_\5 R_\2+R_\1 R_\6 R_\2\cr
&&-18 R_\7 R_\2-6 R_\3^3+6 R_\1 R_\4^2-18 R_\1 R_\3 R_\5-6 R_\4 R_\5-R_\1^3 R_\6\cr
&&+18 R_\3 R_\6+6 R_\1 R_\8
\,,\cr
\widetilde {\cal V}^\9_3 &=& R_\5 R_\1^4-4 R_\6 R_\1^3-6 R_\2^2 R_\3 R_\1^2+6 R_\3 R_\4 R_\1^2+6 R_\2 R_\5 R_\1^2+5 R_\2^4 R_\1\cr
&&+24 R_\2 R_\3^2 R_\1+3 R_\4^2 R_\1-18 R_\2^2 R_\4 R_\1-40 R_\3 R_\5 R_\1+4 R_\2 R_\6 R_\1\cr
&&+18 R_\8 R_\1-12 R_\3^3-14 R_\2^3 R_\3+6 R_\2 R_\3 R_\4+39 R_\2^2 R_\5-6 R_\4 R_\5\cr
&&+36 R_\3 R_\6-48 R_\2 R_\7
\,,\cr
\widetilde {\cal V}^{\smallten}_1 &=& R_\2^5-2 R_\1 R_\3 R_\2^3-8 R_\4 R_\2^3+5 R_\3^2 R_\2^2+6 R_\1 R_\5 R_\2^2+9 R_\6 R_\2^2\cr
&&+R_\1^2 R_\3^2 R_\2+11 R_\4^2 R_\2+2 R_\1 R_\3 R_\4 R_\2-18 R_\3 R_\5 R_\2-R_\1^2 R_\6 R_\2\cr
&&-8 R_\1 R_\7 R_\2-2 R_\8 R_\2-2 R_\1 R_\3^3+8 R_\5^2-2 R_\1^2 R_\3 R_\5-2 R_\1 R_\4 R_\5\cr
&&+6 R_\1 R_\3 R_\6-14 R_\4 R_\6+8 R_\3 R_\7+2 R_\1^2 R_\8
\,,\cr
\widetilde {\cal V}^\smallten_2 &=& 2 R_\2^5-3 R_\1 R_\3 R_\2^3-18 R_\4 R_\2^3+9 R_\3^2 R_\2^2+R_\1^2 R_\4 R_\2^2+7 R_\1 R_\5 R_\2^2\cr
&&+26 R_\6 R_\2^2+26 R_\4^2 R_\2+5 R_\1 R_\3 R_\4 R_\2-32 R_\3 R_\5 R_\2-2 R_\1^2 R_\6 R_\2\cr
&&-10 R_\1 R_\7 R_\2-20 R_\8 R_\2-R_\1^2 R_\4^2+14 R_\5^2-7 R_\3^2 R_\4-3 R_\1 R_\4 R_\5\cr
&&-2 R_\1 R_\3 R_\6-28 R_\4 R_\6+28 R_\3 R_\7+2 R_\1^2 R_\8+6 R_\1 R_\9
\,,\cr
\widetilde {\cal V}^\smallten_3 &=& 3 R_\2^5-4 R_\1 R_\3 R_\2^3-26 R_\4 R_\2^3+12 R_\3^2 R_\2^2+12 R_\1 R_\5 R_\2^2+36 R_\6 R_\2^2\cr
&&+36 R_\4^2 R_\2+9 R_\1 R_\3 R_\4 R_\2-45 R_\3 R_\5 R_\2-21 R_\1 R_\7 R_\2-24 R_\8 R_\2\cr
&&-3 R_\1^2 R_\4^2+18 R_\5^2-10 R_\3^2 R_\4+R_\1^3 R_\3 R_\4-3 R_\1^2 R_\3 R_\5-2 R_\1 R_\3 R_\6\cr
&&-40 R_\4 R_\6-R_\1^3 R_\7+40 R_\3 R_\7+6 R_\1^2 R_\8+6 R_\1 R_\9
\,,\cr
\widetilde {\cal V}^\smallten_4 &=& 5 R_\2^5-6 R_\1 R_\3 R_\2^3-44 R_\4 R_\2^3+18 R_\3^2 R_\2^2+15 R_\1 R_\5 R_\2^2+67 R_\6 R_\2^2\cr
&&+57 R_\4^2 R_\2+18 R_\1 R_\3 R_\4 R_\2+R_\1^3 R_\5 R_\2-67 R_\3 R_\5 R_\2-3 R_\1^2 R_\6 R_\2\cr
&&-27 R_\1 R_\7 R_\2-54 R_\8 R_\2+30 R_\5^2-18 R_\3^2 R_\4-3 R_\1^2 R_\3 R_\5\cr
&&-12 R_\1 R_\4 R_\5-6 R_\1 R_\3 R_\6-58 R_\4 R_\6-R_\1^3 R_\7+64 R_\3 R_\7\cr
&&+6 R_\1^2 R_\8+18 R_\1 R_\9
\,.\label{tvres}
\eea

\section{Radially conserved current and boundary stress tensor}

\label{sec:rccbst}

In this appendix we construct the radially-conserved current for some linear perturbations on the AdS planar black hole background and show that they match the relevant boundary stress tensor.  The matching was demonstrated \cite{Liu:2017kml} for ghost free theories such as Lovelock gravities or a class of Horndeski theories. It was conjectured in \cite{Liu:2017kml} that the matching holds for general gravity theories that may or may not have ghost excitations.  In this appendix, we would like to prove this conjecture for the theory we studied in section \ref{sec:AdS/CFT}. The theory is Einstein gravity extended with cubic Ricci tensor polynomials, studied in section \ref{sec:cubic}.  We shall consider the general case with arbitrary $(e_1,e_2,e_3)$ coupling constants.  For simplicity, we shall consider only the pure gravity theory in four dimensions without any matter.  The theory admits the Schwarzschild AdS planar black hole
\be
ds^2 = - f dt^2 + \fft{dr^2}{f} + r^2 (dx_1^2 + dx_2^2)\,,\qquad
f=g^2 r^2 -\fft{\mu}{r}\,.
\ee
Note that since the matching of the bulk current and boundary stress tensor occurs at the asymptotic infinity of the AdS spacetime, our results are applicable when additional minimally coupled matter is included, as long as the asymptotic structure is maintained.  This is indeed the case for the matter system discussed in section \ref{sec:AdS/CFT}. The equations of motion imply that
\be
\Lambda_0 = -3g^2 (1 - 144 e_1 g^4 - 36 e_2 g^4 - 9 e_3 g^4)\,.
\ee
We consider two types of perturbations, namely
\bea
\hbox{case 1}:&& dx_1^2 + dx_2^2 \rightarrow dx_1^2 + dx_2^2 + 2 \Psi_1\, dx_1 dx_2\,,\\
\hbox{case 2}:&& ds^2 \rightarrow ds^2 + 2 \Psi_2\, dt dx_1\,,
\eea
where
\be
\Psi_1 =- \varsigma\, t + \psi_1(r)\,,\qquad
\Phi_2 = - r^2 \zeta\, t + \psi_2(r)\,.
\ee
The first perturbation is associated with holographic shear viscosity; the second
perturbation is associated with the holographic heat current.  (Note that we consider the two perturbations independently.) The linearized equations of motion for $\psi_1$ and $\psi_2$ are given by
\bea
&&-\ft12 (1 + 432 e_1 g^4) r \big((3 g^2 r^2+f) \psi_1' + r f\psi_1''\big)\cr
&& + \ft32 (4 e_2 + 3 e_3) g^2 r^2 f^2\psi_1''''+18(e_2 + 3e_3) g^4 r^3 f \psi_1'''\cr
&&+ \ft32  g^2 (-2 (4 e_2 + 3 e_3) f^2 +
9 (4 e_2 + 5 e_3) g^2 r^2 f+ 18 (4 e_2 + 3 e_3) g^4 r^4)\psi_1''\cr
&& \ft32 g^2 \big(4 (4 e_2 + 3 e_3) f^2 - 3 (20 e_2 + 9 e_3) f g^2 r^2 +
   9 (-4 e_2 + 3 e_3) g^4 r^4\big) \fft{1}{r} \psi_1'=0\,,\nn\\
&&-\ft12(1 + 432 e_1 g^4)f\, \big(\psi_2'' - \fft{2}{r^2} \psi_2\big) +
\ft32 (4e_2 + 3e_3)g^2 f^2 \psi_2''''\cr
&&+ 3 (4 e_2 + 3 e_3) g^2 f \big(3g^2 r - \fft{f}{r}\big) \psi_2''' -
3g^2 f \big(\ft92 (4e_2 + e_3) g^2 +  (4e_2 +3e_3) \fft{f}{r^2}\big)\psi_2''\cr
&& - 18 (4e_2 + 3 e_3) g^2 f(g^2 r^2 - f) \fft{1}{r^3} \psi_2' \cr
&&+ 3 g^2 f  \big(-10 (4 e_2 + 3 e_3) f + 3 (28 e_2 + 15 e_3) g^2 r^2
\big) \fft{1}{r^4} \psi_2
=0\,.
\eea
These two equations are integrable and give rise to radially conserved currents.  In \cite{Liu:2017kml}, a general formula for such conserved current was obtained using a Noether procedure for a variety of gravities.  To be specific, it was stated that the quantity ${\cal J}^{x_1}$ of
(\ref{jx1}) is radially conserved, where $J^{\mu\nu}$ is given in (\ref{J2grav}).  For the case 1 and case 2 perturbations, the relevant Killing vectors are $\xi=\partial_{x_2}$ and
$\xi=\partial_{t}$ respectively.  For the background metric that is diagonal, ${\cal J}^{x_1}$ vanishes identically and hence conserved. For the two perturbations we consider in this section, the linearized quantities are given by
\bea
16\pi\Big({\cal J}^{x_1}_{1}\Big)^{\rm lin} &=& -(1 + 432 e_1 g^4) r^2 f \psi_1' + 3g^2 f
\Big( (4e_2+3e_3) r^2 (f \psi_1''' + 6g^2 r \psi_1'') \cr
&&\qquad\qquad\qquad-\big(
2 (4 e_2 + 3 e_3) f + 3 (4 e_2 - 3 e_3) g^2 r^2\big)\psi_1'
\Big)\,,\nn\\
16\pi\Big({\cal J}^{x_1}_{2}\Big)^{\rm lin} &=& (1 + 432 e_1 g^4) \big(f\psi_2'+\fft{1}{r}(f-3g^2r^2) \psi_2\big)\\
&& - \fft{3g^2}{r^3}
\Big((4e_2+3e_3) r^3 f^2 \psi_2''' - rf \big(9(4e_2 + e_3) g^2 r^2 +
2(4e_2+3e_3)f\big) \psi_2'\cr&&\qquad+\big(4 (4 e_2 + 3 e_3) f^2 - 9 (4 e_2 + e_3) f g^2 r^2 + 27 (4 e_2 + e_3) g^4 r^4\big)\psi_2\Big)\,.\nn
\eea
It is easy to verify that a radial derivative of the above two quantities both vanish provided that the corresponding equations of motion for $\psi_1$ or $\psi_2$ are satisfied.

We now derive the boundary stress tensor associated with these two perturbations.  In order to derive the boundary stress tensor, it is necessary to construct both the generalized Gibbons-Hawking surface term and the holographic boundary counterterms.  The generalized Gibbons-Hawking terms are given by
\cite{deruelle}
\be
S_{\rm surf} = \fft{1}{8\pi} \int d^{D-1} x\, \sqrt{-h}\, \Phi^\mu_\nu\, K_\mu^\nu\,,\qquad
\Phi^\mu_\nu = 2 P^\mu{}_{\rho\nu\sigma} n^\rho n^\sigma\,.
\ee
Here $K_{\mu\nu}=h_\mu{}^\rho \nabla_\rho n_\nu$ is the second fundamental form
and $h_{\mu\nu}=g_{\mu\nu}-n_\mu n_\nu$, with $n^\mu$ being the unit vector normal to the surface.  For our specific example, we have $n=\partial_r/\sqrt{f}$.  It is important to emphasize that when we vary the surface action $S_{\rm surf}$, we should treat the quantity
$\Phi^\mu_\nu$ as an auxiliary field that is not varied. In other words, for the variation of the action, we have
\be
\delta S_{\rm surf} = \fft{1}{8\pi} \int d^{D-1} x\, \Phi^\mu_\nu\, \delta \Big(\sqrt{-h}
K_\mu^\nu\Big)\,.
\ee
It can then be verified that the surface term plays the precise role of getting rid of the
$\nabla_\rho \delta g_{\mu\nu}$ terms arising from the integration by parts after the variation of the bulk action.

Finally we need to construct the holographic boundary counterterms. Since we consider AdS planar black holes with a flat boundary, there is only one relevant contribution to the boundary counterterm, given by
\be
S_{\rm ct}= \fft{1}{16\pi} \int d^{D-1}x\, \sqrt{-h}\, c\,,\qquad c=
-4g - 108 (16e_1 + 4 e_2 + e_3) g^5\,.
\ee
Note that the constant $c$ is determined by requiring that the total action,
\be
S_{\rm tot} = S_{\rm bulk} + S_{\rm surf} + S_{\rm ct}\,,\label{Stot}
\ee
is finite for the AdS vacuum.  Having obtained the full action, we can derive the
boundary stress tensor at the linear level, given by
\bea
16\pi T^{x_2 x_1}_{\rm lin} &=& \fft{2}{\sqrt{-h}} \fft{\delta S_{\rm tot}}{\delta h_{x_2x_1}} =  \fft{2}{\sqrt{-h}} \fft{\delta S_{\rm tot}}{\delta \psi_1}\cr
&=& -\ft{1 + 432 e_1 g^4}{g r^2}\Big( \big(-4 \sqrt{f} g r+f+3 g^2 r^2\big) \psi _1+r f \psi _1'\Big)\cr
&&+ \ft{3 \left(4 e_2+3 e_3\right) g f^2}{r} \psi_1''' + 18 \left(4 e_2+3 e_3\right)  g^3 f \psi_1 ''\cr
 &&-\ft{3 g f \left(\left(8 e_2+6 e_3\right) f+3 \left(4 e_2-3 e_3\right) g^2 r^2\right)}{r^3}\psi_1 '\cr
&&-\ft{27 \left(4 e_2+e_3\right) g^3 \left(-4 g r\sqrt{f}+f+3 g^2 r^2\right)}{r^2}
\psi_1\,,\nn\\
16\pi T^{t x_1}_{\rm lin} &=& \fft{2}{\sqrt{-h}} \fft{\delta S_{\rm tot}}{\delta h_{tx_1}} =  \fft{2}{\sqrt{-h}} \fft{\delta S_{\rm tot}}{\delta \psi_2}\cr
&=&\ft{1 + 432 e_1 g^4}{g r^4}
\Big(\big(2-\ft{4 g r}{\sqrt{f}}\big)\psi_2+r \psi_2 '\Big) -\ft{3 \left(4 e_2+3 e_3\right) g f }{r^3}\psi_2'''\cr
&&-\ft{9 \left(4 e_2+3 e_3\right) g  \left(g^2 r^2-f\right)}{2 r^4}\psi_2 ''+\ft{3 g  \left(\left(8 e_2+6 e_3\right) f+9 \left(4 e_2+e_3\right) g^2 r^2\right)}{r^5}\psi_2 '\cr
&&-\ft{3 g \left(7 \left(4 e_2+3 e_3\right) f^{3/2}-3 \left(28 e_2+9 e_3\right) \sqrt{f} g^2 r^2+36 \left(4 e_2+e_3\right) g^3 r^3\right)}{\sqrt{f} r^6} \psi_2 \,.
\eea
It is then straightforward to verify that
\bea
\fft{{\cal J}_{1,\rm lin}^{x_1}}{g r^3\, T_{\rm lin}^{x_2x_1}}\Big|_{r\rightarrow \infty} = 1\,,\qquad \fft{{\cal J}_{2,\rm lin}^{x_1}}{g^3 r^5\, T_{\rm lin}^{tx_1}}\Big|_{r\rightarrow \infty} = 1
\eea
Thus we see that the radially conserved currents proposed in \cite{Liu:2017kml} match precisely the corresponding stress tensor components in this cubic Ricci polynomial gravity for generic coupling constants $(e_1,e_2,e_3)$.  When $(e_1,e_2,e_3)$ are given by (\ref{Q31}), the contributions from the cubic terms to the ${\cal J}_2^{x_1}$ and $T^{tx_1}$, associated with the holographic heat current, vanish identically.

            The proof of the matching of the radially conserved bulk current with the associated
boundary stress tensor is not as elegant and covariant as that in \cite{Liu:2017kml} for ghost free theories such as Lovelock gravities.  Our concrete demonstration for Ricci cubic gravities nevertheless serves as a strong evidence that the proposed radially conserved current of \cite{Liu:2017kml} is indeed the bulk dual of the holographic heat current.

\end{document}